\documentclass[aps,reprint,nofootinbib,onecolumn,notitlepage,preprintnumbers]{revtex4-1}
\pdfoutput=1

\usepackage{amsmath}
\usepackage{amssymb}
\usepackage{siunitx}
\usepackage{bbm}
\usepackage{booktabs}
\usepackage{braket}
\usepackage{diagbox}
\usepackage{graphicx}
\usepackage{hyperref}
\usepackage[utf8]{inputenc}
\usepackage{multirow}
\usepackage{placeins}
\usepackage{slashed}
\usepackage[dvipsnames]{xcolor}
\usepackage{xspace}

\newcommand{\eps}{\varepsilon}
\newcommand{\refapp}[1]{appendix~\ref{app:#1}}

\newcommand{\reffig}[1]{figure~\ref{fig:#1}}
\newcommand{\refsec}[1]{section~\ref{sec:#1}}
\newcommand{\reftab}[1]{table~\ref{tab:#1}}

\newcommand{\order}[1]{\mathcal{O}\left(#1\right)}

\newcommand{\GeV}{\si{\giga\electronvolt}}

\newcommand{\EOS}{\texttt{EOS}\xspace}

\AtBeginDocument{
  \heavyrulewidth=.08em
  \lightrulewidth=.05em
  \cmidrulewidth=.03em
  \belowrulesep=.65ex
  \belowbottomsep=0pt
  \aboverulesep=.4ex
  \abovetopsep=0pt
  \cmidrulesep=\doublerulesep
  \cmidrulekern=.5em
  \defaultaddspace=.5em
}

\begin{document}

\title{Heavy-Quark Expansion for $\boldsymbol{\bar{B}_s\to D^{(*)}_s}$ Form Factors and Unitarity Bounds beyond the $\boldsymbol{SU(3)_F}$ Limit}

\author{Marzia Bordone}
\email{marzia.bordone@uni-siegen.de}
\affiliation{Universit\"at Siegen, Walter-Flex Stra\ss{}e 3, 57072 Siegen, Germany}

\author{Nico Gubernari}
\email{nicogubernari@gmail.com}
\affiliation{Technische Universit\"at M\"unchen, James-Franck-Stra\ss{}e 1, 85748 Garching, Germany}

\author{Martin Jung}
\email{martin.jung@unito.it}
\affiliation{Dipartimento di Fisica, Universit\`a di Torino \& INFN, Sezione di Torino, I-10125 Torino, Italy }

\author{Danny van Dyk}
\email{danny.van.dyk@gmail.com}
\affiliation{Technische Universit\"at M\"unchen, James-Franck-Stra\ss{}e 1, 85748 Garching, Germany}

\begin{abstract}
We carry out a comprehensive analysis of the full set of $\bar{B}_q \to D_q^{(*)}$ form factors for spectator quarks $q=u,d,s$
within the framework of the Heavy-Quark Expansion (HQE) to order $\order{\alpha_s, 1/m_b, 1/m_c^2}$. In addition to the available lattice QCD calculations we make
use of two new sets of theoretical constraints: we produce for the first time numerical predictions for the full set of
$\bar{B}_s \to D_s^{(*)}$ form factors using Light-Cone Sum Rules with $B_s$-meson distribution amplitudes. Furthermore, we reassess
the QCD three-point sum rule results for the Isgur-Wise functions entering all our form factors for both $q=u,d$ and
$q=s$ spectator quarks.
These additional constraints allow us to go beyond the commonly used assumption of $SU(3)_F$ symmetry for the $\bar B_s\to D_s^{(*)}$ form factors, especially in the unitarity constraints which we impose throughout our analysis.
We find the coefficients of the IW functions emerging at $\order{1/m_c^2}$ to be consistent with the naive $\order{1}$ expectation, indicating a good convergence of the HQE.
While we do not find significant $SU(3)$ breaking, the explicit treatment of $q=s$ as compared to a simple symmetry assumption renders the unitarity constraints more effective.
We find that the (pseudo)scalar bounds are saturated to a large degree, which affects our theory predictions. 
We analyze the phenomenological consequences of our improved form factors by extracting $|V_{cb}|$ from $\bar B\to D^{(*)}\ell\nu$ decays and producing theoretical predictions for the lepton-flavour universality ratios $R(D)$, $R(D^*)$, $R(D_s)$ and $R(D_s^*)$, as well as the $\tau$- and $D_q^*$ polarization fractions for  the $\bar B_q\to D_q^{(*)}\tau\nu$ modes.
\end{abstract}

\preprint{EOS-2019-04, P3H-19-050, SI-HEP-2019-20, TUM-HEP 1241/19}

\maketitle

\section{Introduction}

Semileptonic $b\to c$ transitions are of great phenomenological interest, both within the Standard Model (SM) and beyond. The foreseeable improved precision for the corresponding measurements \cite{Kou:2018nap,Cerri:2018ypt,Bediaga:2012py} requires a corresponding improvement in their theoretical description. Additional interest is created by the long-standing tensions in semitauonic decays and neutral-current $b\to s \ell^+\ell^-$ transitions, as well as the difference between inclusive and exclusive determinations of the CKM matrix element $|V_{cb}|$ \cite{Amhis:2019ckw}. These tensions motivate potential new physics (NP) contributions to $b\to c\ell\nu$ transitions with light leptons, which in turn require a determination of the corresponding form factors independent of the experimental input. This situation, together with the recent appearance of several experimental analysis allowing for a model-independent interpretation of their results on $\bar B\to D^{(*)}\ell\nu$ decays \cite{Glattauer:2015teq,Abdesselam:2017kjf,Abdesselam:2018nnh}
sparked renewed interest in the relevant hadronic matrix elements \cite{Bernlochner:2017jka,Bigi:2016mdz,Bigi:2017jbd,Bigi:2017njr,Grinstein:2017nlq,Bernlochner:2017xyx,Jaiswal:2017rve,Jung:2018lfu,Gambino:2019sif,Cohen:2019zev,Das:2019cpt}.
A recent theory analysis of the form factors parametrizing the $\bar{B}\to D^{(*)}$ matrix element by three of us \cite{Bordone:2019vic}
uses the heavy-quark expansion to determine the full set of relevant parameters up to order $1/m_c^2$ for the first time, building and improving on the work presented in refs.~\cite{Caprini:1997mu,Bernlochner:2017jka,Jung:2018lfu} in particular.
This analysis also uses unitarity bounds to restrict the parameter space of the so-called Isgur-Wise (IW)
functions \cite{Isgur:1989vq,Isgur:1989ed}. Specifically, it was observed that for $J^P=0^+$ and $0^-$ currents the present results
saturate the bounds to a large degree.
This observation triggers our interest, since $SU(3)_F$ symmetry breaking at the level of $20\%$ is assumed for the form factors in these bounds, with the effect of \emph{lowering} the values of the form factors and therefore
lowering the contributions to the unitarity bounds by $40\%$. For a precision analysis of
the form factors this assumption should be removed, and instead a simultaneous analysis
of the form factors for both light ($q=u,d$) and strange ($q=s$) spectator quarks is warranted.
The purpose of this article is to carry out such a simultaneous analysis. For this work
we take the following steps:
\begin{itemize}
    \item We estimate the normalization and the slope of the subleading IW functions $\eta$, $\chi_{2}$,
    and $\chi_3$, as well as $\eta^{(s)}$, $\chi_{2}^{(s)}$, and $\chi_3^{(s)}$ in the point of zero hadronic recoil,
    based on existing analytical formulas in the literature \cite{Neubert:1992wq,Neubert:1992pn,Ligeti:1993hw}.
    The necessary numerical inputs and our numerical results are compiled in \refapp{QCDSRs}.

    \item We estimate the full set of $\bar{B}_s \to D_s^{(*)}$ form factors needed for the basis of dimension-six
    effective operators $b\to c\ell\nu$ using light-cone sum rules with $B$-meson
    LCDAs. The relevant analytical results have been recently published in ref.~\cite{Gubernari:2018wyi}, and their
    numerical implementation as part of the \EOS software \cite{EOS} facilitates this step. The necessary
    numerical inputs and our numerical results including correlation information are compiled in \refapp{LCSRs}.
    Our numerical results for the $\bar{B}_s\to D_s^*$ transitions allow to carry out our analysis at the complete
    $\order{1/m_c^2}$ level.

    \item With the theoretical constraints at hand, we simultaneously infer the parameters of the various form factors
    within the HQE, in three different fit models. In all of our analyses, we impose the strong unitarity bounds for \emph{all}
    $\bar{B}_q^{(*)} \to D_q^{(*)}$ transitions.
\end{itemize}
Note that some of the results from ref.~\cite{Bordone:2019vic} are superseded by our new results.\\

The structure of this article is as follows. We briefly introduce the necessary notation
and set up our analysis in \refsec{notation}. We discuss the results in \refsec{results}, and
summarize in \refsec{summary}.
In \refapp{QCDSRs} we provide the numerical inputs and results of the updated QCDSR analysis.
In \refapp{LCSRs} we provide the numerical inputs and results of our LCSR analysis.

\section{Notation and Setup}
\label{sec:notation}

We analyse the full set of hadronic matrix elements for the basis of local dimension-three currents $\bar{c}\,\Gamma b$ in $\bar{B}_q \to D_q^{(*)}$
transitions.
For this purpose, we use the heavy-quark expansion (HQE) as reviewed in ref.~\cite{Bernlochner:2017jka}, and
as applied recently in refs.~\cite{Jung:2018lfu,Bordone:2019vic}. Within the expansion of a generic form factor $h(w)$,
\begin{equation}
    h(w) = \xi(w)\hat h(w) = \xi(w)\left(
        a + \hat\alpha_s b
        + \eps_b \, c_b^{(i)}\left[\hat L_i(w)\right]
        + \eps_c \, c_c^{(i)}\left[\hat L_i(w)\right]
        + \eps_c^2 \, d^{(i)}\left[\hat\ell_i(w)\right]
    \right)\,
\end{equation}
one encounters three expansion parameters: $\hat\alpha_s \equiv \alpha_s/\pi$, $\eps_b \equiv \bar\Lambda/(2 m_b)$, and $\eps_c \equiv \bar\Lambda/(2 m_c)$.
The coefficients $a$, $b$, $c_b^{(i)}$, $c_c^{(i)}$, and $d^{(i)}$ in this expansion are linear combinations of Wilson coefficients from the matching of HQET onto QCD
and kinematic functions. The objects $\xi(w)$, $L_i(w)$, and $\ell_i(w)$ are matrix elements of the effective operator in HQET, the IW functions.
To differentiate between matrix elements with a light or a strange spectator quark,
we will add the label ``$(s)$'' where appropriate. This includes the IW functions entering $\bar{B}_s^{(*)}\to D_s^{(*)}$,
as well as $\bar{\Lambda}_{(s)}$, the energy of the light degrees of freedom within the heavy meson in the heavy-quark limit.
For the analysis at hand we use the same power counting of the HQE as introduced in ref.~\cite{Bordone:2019vic}
(i.e., $\hat\alpha_s \sim \eps_b \sim \eps_c^2 \sim \eps^2$).
We also use the same nominal fit model, \emph{i.e.}, the $3/2/1$ model, where the digits refer to the power in the $z$ expansion to which the IW functions
are expanded at different orders in $1/m_q$. In case of the $3/2/1$ model we use $z^3$ for the leading IW function, $z^2$ for subleading IW functions, and $z^1$ for the subsubleading IW functions.

A key point of our analysis is the treatment of the $SU(3)_F$ symmetry breaking in all IW functions. For a large part of the analysis
we do not make any assumption about the size of this breaking, but simply parametrize the $q=s$ IW functions with independent parameters.
Only in one of our scenarios, to be discussed below, we assume that the subsubleading IW functions behave schematically as
\begin{equation}
    \ell_i^{(s)}(w) = \ell_i(w)+\eps_F \delta_{\ell_i}(w)\,,
\end{equation}
with $\eps_F\sim\eps_c$ (and $\delta_{\ell_i}\sim\ell_i$). This assumption is subsequently confronted with the available data.

In addition to the theoretical constraints for the $\bar{B}\to D^{(*)}$ form factors with spectators $q=u,d$ as used in ref.~\cite{Bordone:2019vic},
we include further theory information on the form factors with $q=s$ spectators.
The individual \emph{changed or new} pieces of theory information entering the likelihood are:
\begin{description}
    \item[Lattice] For $\bar{B}_s\to D_s$ the HPQCD collaboration~\cite{McLean:2019qcx} has determined both the vector form factor $f_+^{(s)}$
    and the scalar form factor $f_0^{(s)}$ at non-zero hadronic recoil $w\geq 1$.
    Accounting for the fact that at $w = w_{\text{max},D}$ the two form factors fulfill an equation of motion,
    we can produce $5$ correlated pseudo data points from the correlated parameters provided in ref.~\cite{McLean:2019qcx}.
    In addition, lattice QCD data by the ETM collaboration has been used in ref.~\cite{Atoui:2013zza} to determine $f_+^{(s)}$
    and  the ratios $f_T^{(s)}/f_+^{(s)}$ and $f_0^{(s)}/f_+^{(s)}$ close to the zero recoil point at $q^2 = 11.5\,\GeV^2$. We do not use the
    the result for $f_+^{(s)}$ nor for the ratio $f_0^{(s)}/f_+^{(s)}$ due to their large uncertainties, which are not competitive with
    the HPQCD result. We do use, however, the ratio $f_T^{(s)}/f_+^{(s)}$, thereby including one further data point in the 
    fit.\\
    
    In addition to the $q=s$ constraints, ref.~\cite{Atoui:2013zza} also provides results for $\bar{B}\to D$ form factors,
    which did not enter the previous analysis. Following the same argument as before, we only use the constraint on $f_T/f_+$,
    thereby including one further data point.\\
    
    For theory predictions of observables in nonleptonic $\bar{B}\to D\pi$ decays in the context of QCD factorization, the FNAL/MILC collaborations
    have calculated, amongst other quantities, the ratio $f_0^{(s)}(q^2 = M_\pi^2)/f_0(q^2 = M_\pi^2)$~\cite{Bailey:2012rr}.
    The results of ref.~\cite{Bailey:2012rr} for $q=u,d$ form factors are superseded by those of ref.~\cite{Lattice:2015rga},
    which feature much smaller statistical uncertainties.
    This is due to the use of a larger number of gauge ensembles (14 vs 4), which in turn reduces the statistical correlation
    between the results of refs.~\cite{Bailey:2012rr} and \cite{Lattice:2015rga}. We use this $q=s$ vs $q=u,d$ ratio as part
    of our fit whenever both $q=u,d$ \emph{and} $q=s$ information is included.\\
    
    For $\bar{B}_s \to D_s^*$ the HPQCD collaboration has determined the form factor $h_{A_1}^{(s)}$ at zero recoil
    or equivalently $w=1$~\cite{McLean:2019sds}, which is used in all fits.

    \item[QCDSR] The QCD three-point sum rules used to estimate the normalization and slopes for the subleading-power
    IW functions at $w=1$ and for $q=u,d$ can be adapted for the $q=s$ case. In order to ensure a consistent treatment of both
    parametric and systematic uncertainties, we update the $q=u,d$ analysis and carry out a new $q=s$ analysis.
    The relevant inputs, the procedure to determine the uncertainties, and our numerical results are
    discussed in \refapp{QCDSRs}.\\
    
    \item[LCSR] At $w \geq 1.5$ the $\bar{B}_s \to D_s^{(*)}$ form factor are accessible in LCSRs with $B_s$-meson Light-Cone Distribution
    Amplitudes (LCDAs). Following the recent analytic results for the light-cone OPE of the correlation functions underlying
    these sum rules in ref.~\cite{Gubernari:2018wyi}, we produce the first numerical estimates of the $\bar{B}_s\to D_s^{(*)}$ form factors
    in this approach.
    Our numerical results for the $\bar{B}_s \to D_s^{*}$ form factors, including their correlations, are provided as machine-readable
    ancillary files with the arXiv preprint version of this paper. As in ref.~\cite{Gubernari:2018wyi}, we are unable to
    produce reliable results for the $\bar{B}_s \to D_s$ form factor $f_T$, which is therefore not used in our analysis.
    The new constraints contribute a total of $33$ data points to all fits.
    The relevant inputs and the procedure to determine the uncertainties are discussed in \refapp{LCSRs}.
\end{description}

For the fits to the available theory constraints we consider the following scenarios:
\begin{description}
    \item[A] We fit to only $\bar{B}_s\to D_s^{(*)}$ information in the $3/2/1$ model. In order to use the
        unitarity bounds, we multiply the contributions to the bound by a factor $n_s = 2.2$. This factor
        accounts for the strange-spectator contributions, and uses $SU(3)_F$-symmetry to approximate the
        $q=u,d$ spectator contributions. In this way, we allow for symmetry breaking of $-20\%$ on the amplitude level.
        This scenario encompasses $23$ parameters.
    \item[B] We fit simultaneously to $\bar{B}_{u,d,s}\to D_{u,d,s}^{(*)}$ information in the $3/2/1$ model.
        We treat all IW functions for the light and the strange spectator quarks as fully independent. In this way,
        we introduce the least possible amount of correlation between the form factors, which only arises from the
        contributions to the strong unitarity bounds. As a consequence, our fit has twice the number of free parameters
        as in the fit in ref.~\cite{Bordone:2019vic}, corresponding to $46$ parameters.
    \item[C] As scenario B, but additionally we consider the impact of finite $SU(3)_F$ symmetry breaking in the form factors
        by amending our previous power counting: we count the expansion parameter $\eps_F$ for the symmetry breaking
        as $\eps_F \sim \eps \sim \eps_c$. With this power-counting, a generic form factor $h^{(s)}$ receives
        contributions from the subsubleading IW functions that can be expressed schematically as:
        \begin{equation}
        \begin{aligned}
            h^{(s)}(w)
                & \supseteq \eps_c^2 \ell_i^{(s)}(w) = \eps_c^2 \ell_i(w) + \eps_c^2\, \eps_F \delta_{\ell_i}(w)\\
                & \sim \eps^2 \ell_i(w) + \eps^3 \delta_{\ell_i}(w)\,.
        \end{aligned}
        \end{equation}
        Since we discard terms at order $\eps^3$, we suppress the symmetry-breaking terms $\delta_{\ell_i}$ of the
        subsubleading IW functions and identify $\ell_i^{(s)}(w) = \ell_i(w)$ \emph{in this scenario only}.
        In this way, we reduce the number of free parameters to $34$.
\end{description}

\section{Results}
\label{sec:results}

\begin{table}[t!]
    \renewcommand{\arraystretch}{1.1}
    \begin{tabular}{r c c c c c}
        \toprule
        ~                              &
            ~                          &
            A                          &
            B                          &
            C                          &
            C                          \\
        likelihood                     &
            \diagbox{\#data pts}{\#par}     &
            $23$                       &
            $46$                       &
            $34$                       &
            $34$                       \\
        \midrule
        lattice($D_s$) &
            $6$      &
            $ 0.26$  &
            $ 0.49$  &
            $ 1.10$  &
            $ 1.25$  \\
        lattice($D_s^*$) &
            $1$      &
            $ 0.00$  &
            $ 0.00$  &
            $ 0.47$  &
            $ 0.45$  \\
        QCDSR $\bar{B}_s\to D_s^{(*)}$        &
            $5$      &
            $ 0.15$  &
            $ 0.11$  &
            $ 0.29$  &
            $ 0.23$  \\
        LCSR $\bar{B}_s\to D_s^{(*)}$          &
            $33$     &
            $ 1.26$  &
            $ 1.11$  &
            $ 1.43$  &
            $ 1.58$  \\
        \midrule
        lattice($D$) &
            $(13)$   &
            ---      &
            $ 7.16$  &
            $ 8.02$  &
            $ 8.19$  \\
        lattice($D^*$) &
            $(1)$    &
            ---      &
            $ 0.01$  &
            $ 0.59$  &
            $ 0.61$  \\
        QCDSR $\bar{B}\to D^{(*)}$        &
            $(5)$    &
            ---      &
            $ 0.08$  &
            $ 0.24$  &
            $ 0.25$  \\
        LCSR $\bar{B}\to D^{(*)}$         &
            $(33)$   &
            ---      &
            $ 3.11$  &
            $ 2.86$  &
            $ 2.78$  \\
        \midrule
        lattice ratio ($D_s/D$) &
            $(1)$    &
            ---      &
            $ 0.58$  &
            $ 0.77$  &
            $ 1.00$  \\
        \midrule
        $\bar{B}\to D \lbrace e^-,\mu^-\rbrace \bar\nu$ &
            $(9)$    &
            ---      &
            ---      &
            ---      &
            $ 6.87$  \\
        $\bar{B}\to D^* \lbrace e^-,\mu^-\rbrace \bar\nu$ 2017 &
            $(9)$    &
            ---      &
            ---      &
            ---      &
            $ 7.73$  \\
        $\bar{B}\to D^* \lbrace e^-,\mu^-\rbrace \bar\nu$ 2018 &
            $(9)$    &
            ---      &
            ---      &
            ---      &
            $ 5.34$  \\
        \midrule
        \multirow{3}{*}{total} &
            $45$               &
            $ 1.68$            &
            ---                &
            ---                &
            ---                \\
        ~                      &
            $(98)$             &
            ---                &
            $ 12.67$           &
            $ 15.75$           &
            ---                \\
        ~                      &
            $(125)$            &
            ---                &
            ---                &
            ---                &
            $ 36.27$            \\
        \bottomrule
    \end{tabular}
    \renewcommand{\arraystretch}{1}
    \caption{%
        Summary of the goodness of fit in terms of the $\chi^2$ values at the best-fit point for all combinations of fit scenarios and datasets.
    }
    \label{tab:gof}
\end{table}

In the following we describe the fits that are part of our analysis, in terms of the combination of
likelihoods and parameter scenarios, focusing in particular on the impact of the improved unitarity constraints
and quantifying the amount of $SU(3)_F$ breaking. 
A summary of their goodness of fit, expressed through the $\chi^2$
values at the best-fit point, is given in \reftab{gof}.\\

We begin with fitting the restricted theory likelihood, which includes exclusively the 45 $q=s$ data points.
For this likelihood, only fit scenario A with its 23 parameters is applicable. We obtain a very good fit
with a minimal $\chi^2 / \text{d.o.f.} \sim 2 / 22$. The very small $\chi^2$ value is not very surprising, given the large systematic
uncertainties assigned to the QCDSR and LCSR data points.
%The best-fit point obtained here for $q=s$ is \emph{not}
%contained within the $68\%$ probability intervals for the $q=u,d$ parameters obtained in ref.~\cite{Bordone:2019vic}.
A first step to test the assumption of exact $SU(3)_F$ symmetry is taken by evaluating our nominal likelihood at the best-fit point obtained
in ref.~\cite{Bordone:2019vic} from the combination of the theoretical and experimental likelihoods. We find excellent compatibility of the total
likelihood, with an increase of $\chi^2$ at the
$q=u,d$ best-fit point by approximately $15$. Inspecting the individual constraints, we find the largest single increase of $\sim 11$ is caused
by the very precise lattice constraints on the $\bar{B}_s\to D_s$ form factors by the HPQCD collaboration. Nevertheless, even this shift still indicates
reasonable compatibility of the lattice QCD constraint with the central value from ref.~\cite{Bordone:2019vic}, with an individual $p$ value of $\sim 10\%$.
The remaining constraints are perfectly compatible with the assumption of exact $SU(3)_F$ symmetry in this simple comparison.\\

We continue with fitting the nominal theory likelihood, which includes all $q=u,d$ and $q=s$ data points as well
as the single $q=s/q=u,d$ lattice ratio. For this likelihood, only scenarios B and C are applicable with their
$46$ and $34$ parameters, respectively. For scenario B we find an excellent fit with $\chi^2 / \text{d.o.f.} \sim 13 / 52$.
Relative to this result, scenario C increases the d.o.f.~by 12 while only increasing the $\chi^2$ by $\sim 3$.
For both scenarios, the best-fit values of $q=u,d$ parameters are contained within the $68\%$ probability intervals obtained
in ref.~\cite{Bordone:2019vic}. We continue with a model comparison between the fits to the nominal likelihood
in scenarios B and C. Using posterior samples we compute the model evidence for both scenarios,
and hence their Bayes factor:
\begin{equation}
    \log_{10} \frac{P(\text{nominal likelihood}\,|\,\text{scenario B})}{P(\text{nominal likelihood}\,|\,\text{scenario C})}
        = \log_{10} \frac{9.80 \cdot 10^{48}}{2.71 \cdot 10^{57}} \sim -8\,.
\end{equation}
Using Jeffrey's scale for the interpretation of the Bayes factor \cite{Jeffreys:1961}, this result indicates that the nominal likelihood
favours scenario C over scenario B \emph{decisively}, leading to the conclusion that scenario C is --- on average --- much
more efficient than scenario B in describing the data. Hence, we will not use scenario B from this point on.\\

We furthermore tested the compatibility of the theory data with exact $SU(3)_F$ symmetry for \emph{all} parameters in the IW functions. This fit shows a further increase of $\chi^2$ by $\sim 3.5$, to be compared to an increase of the d.o.f. by 11. The data therefore show no indication of $SU(3)_F$ breaking at the present level of precision. We nevertheless refrain from using this limit beyond the subsubleading IW functions, in order to allow for the possibility of a sizable breaking and to include the resulting uncertainty for the form factors and observables.

We finally fit the combined likelihood comprised of the nominal theory likelihood and the experimental likelihood containing
the Belle results of the kinematical PDFs in the recoil variable $w$. Following the model comparison above, we only fit this
likelihood with scenario C. The $\chi^2 / \text{d.o.f.} \sim 36 / 91$ indicates an excellent fit, with an increase of $\chi^2$ by $\sim 20$ for 27 additional d.o.f. Compared to the
fits of the nominal theory likelihood in scenario C, the increase in the minimal $\chi^2$ can be attributed in full
to the experimental likelihood, indicating that either likelihood is well described by scenario C. We provide the best-fit
point as well as the individual one-dimensional $68\%$ probability intervals for each fit parameters in \reftab{BFP}.\\

\begin{figure}[p]
    \begin{tabular}{ccc}
        \includegraphics[width=.35\textwidth]{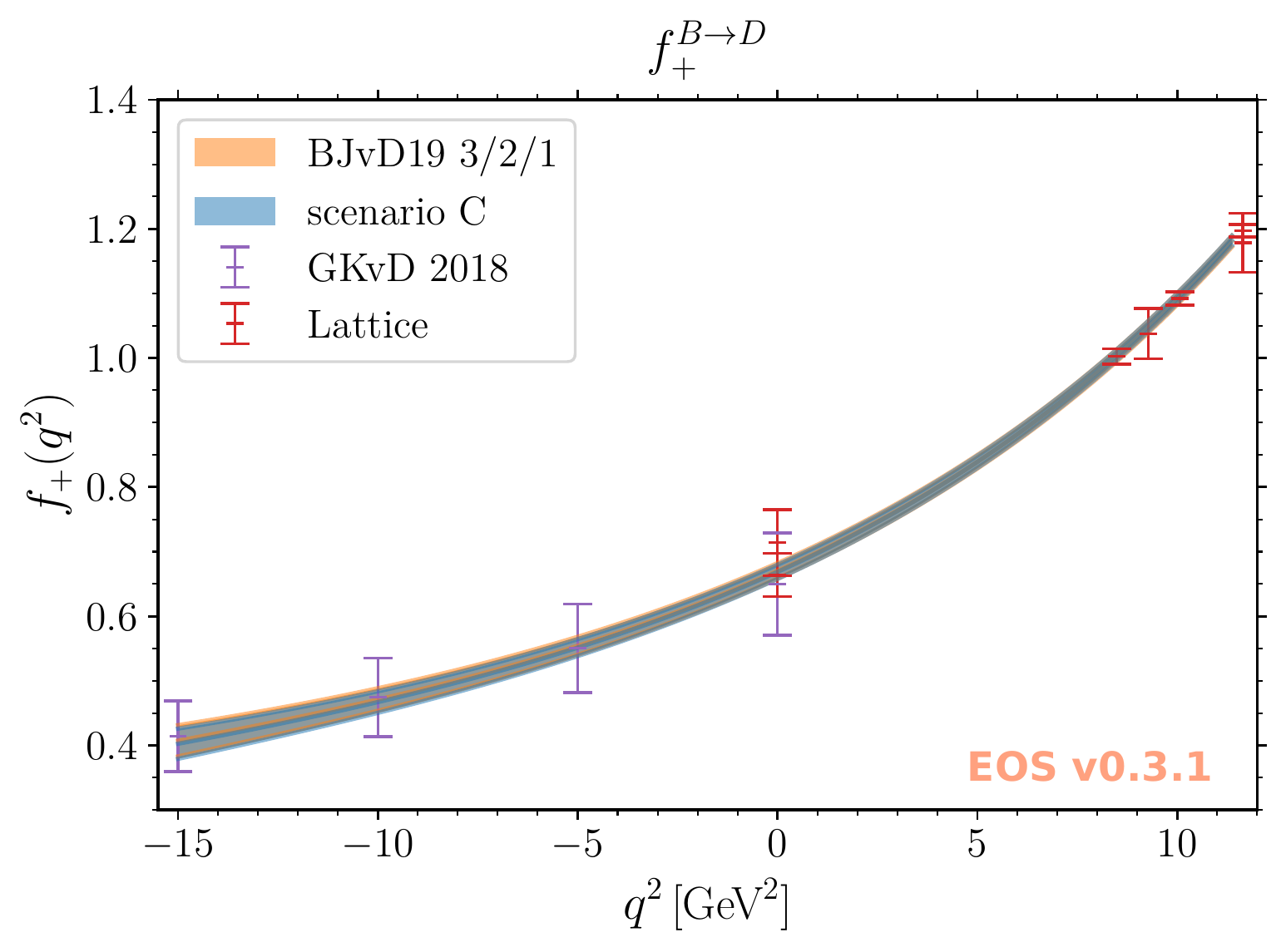}  &
        \includegraphics[width=.35\textwidth]{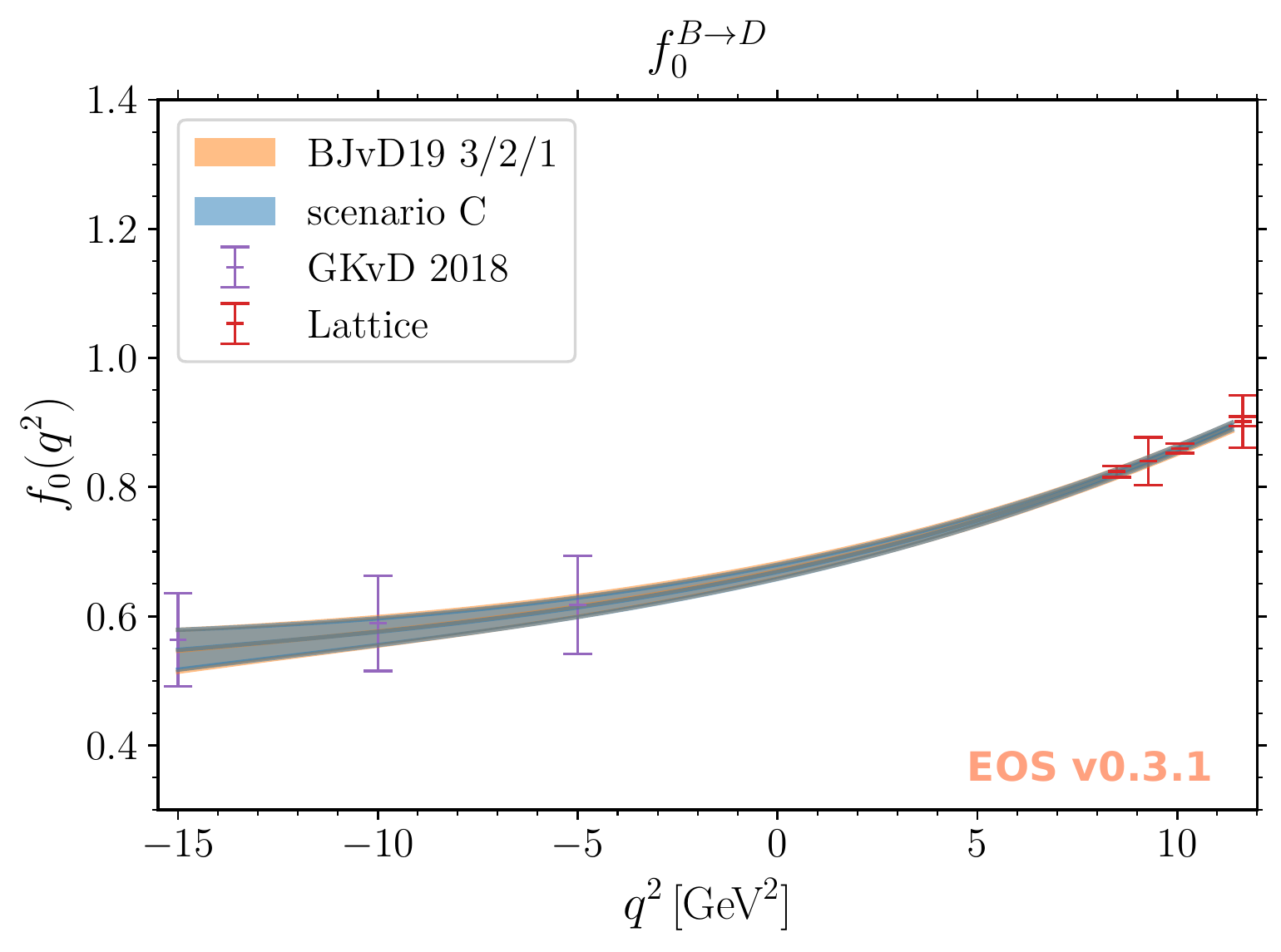}  &
        \includegraphics[width=.35\textwidth]{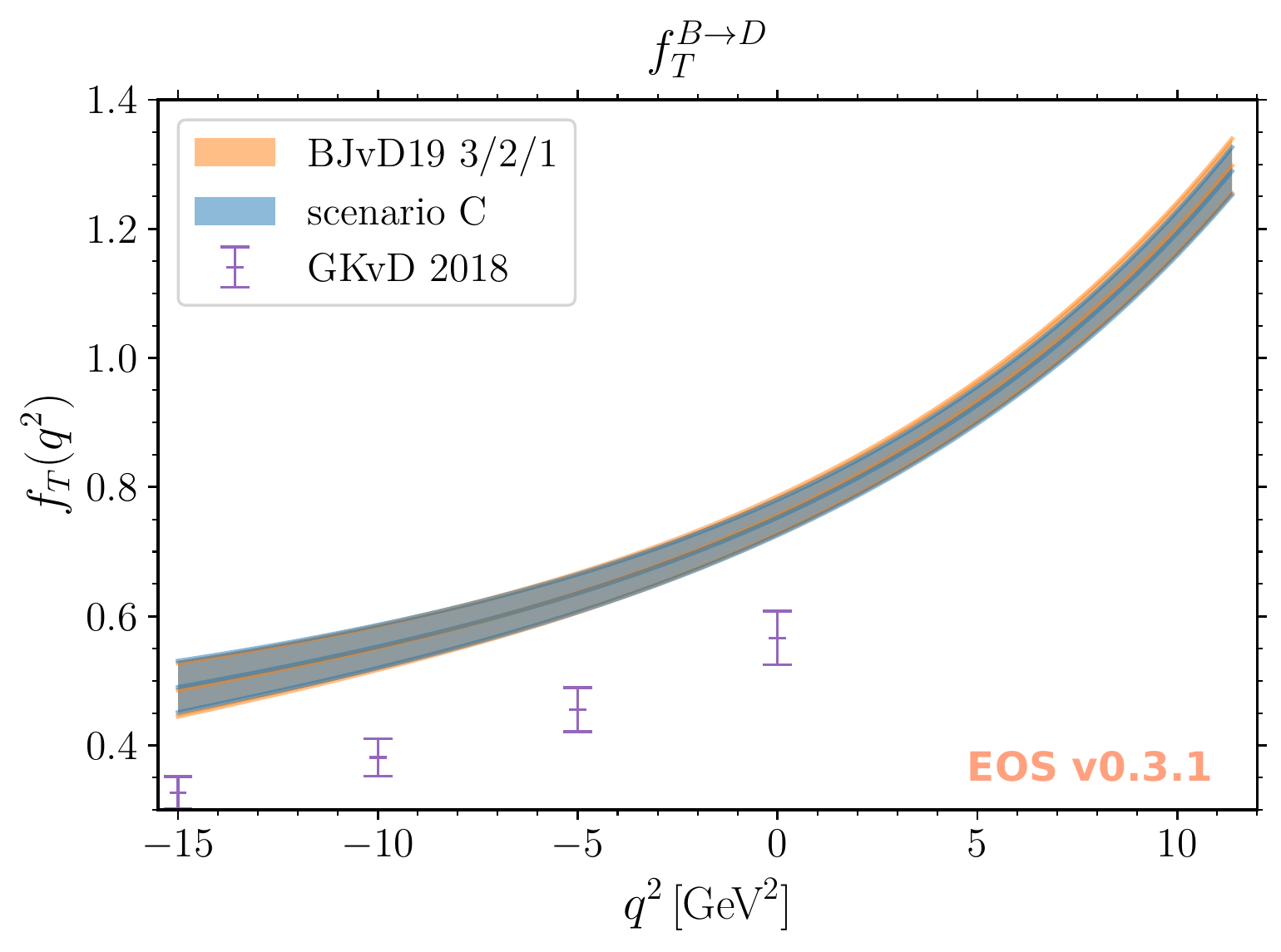}  \\[1.25em]
        \includegraphics[width=.35\textwidth]{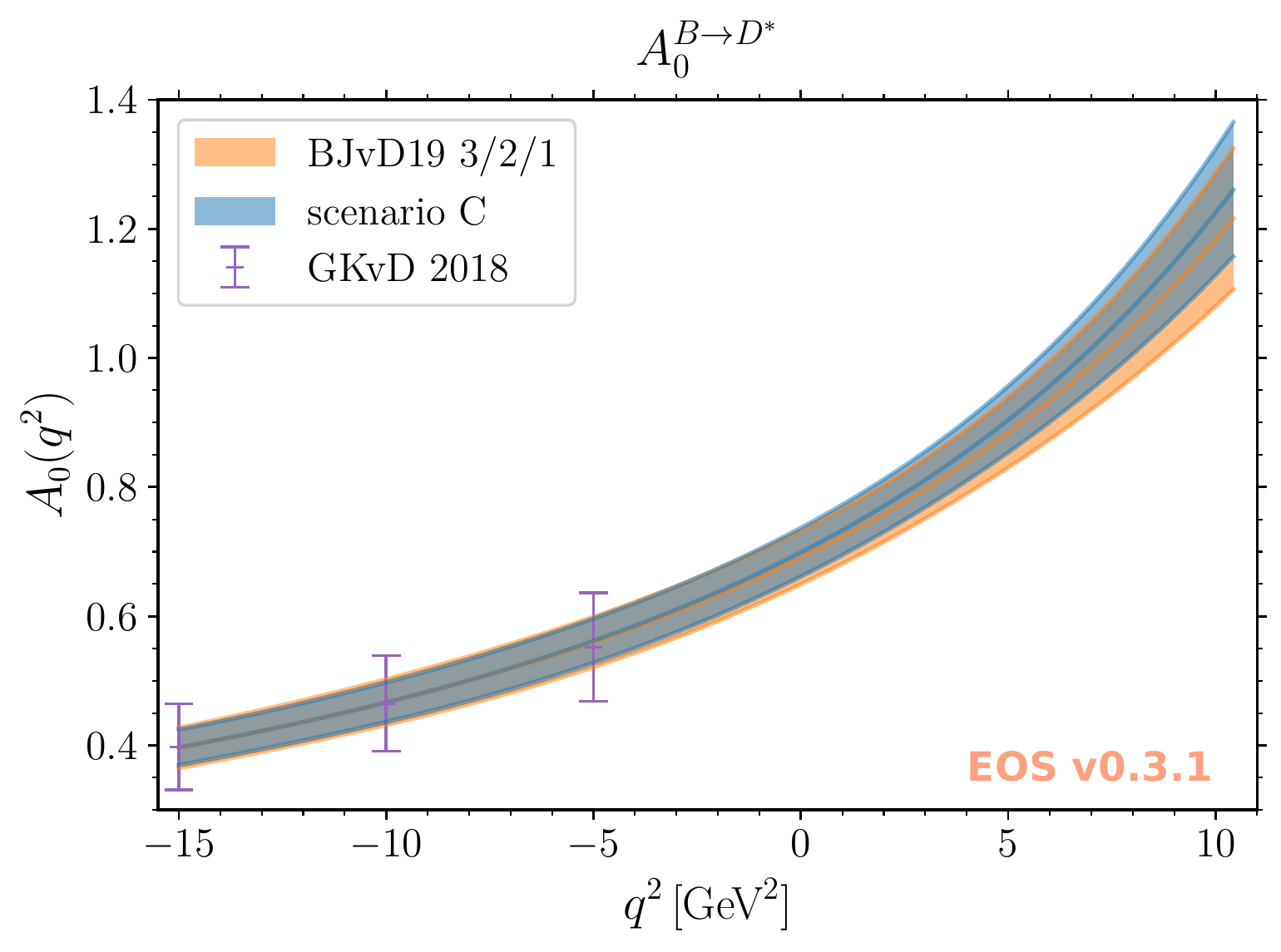}  &
        \includegraphics[width=.35\textwidth]{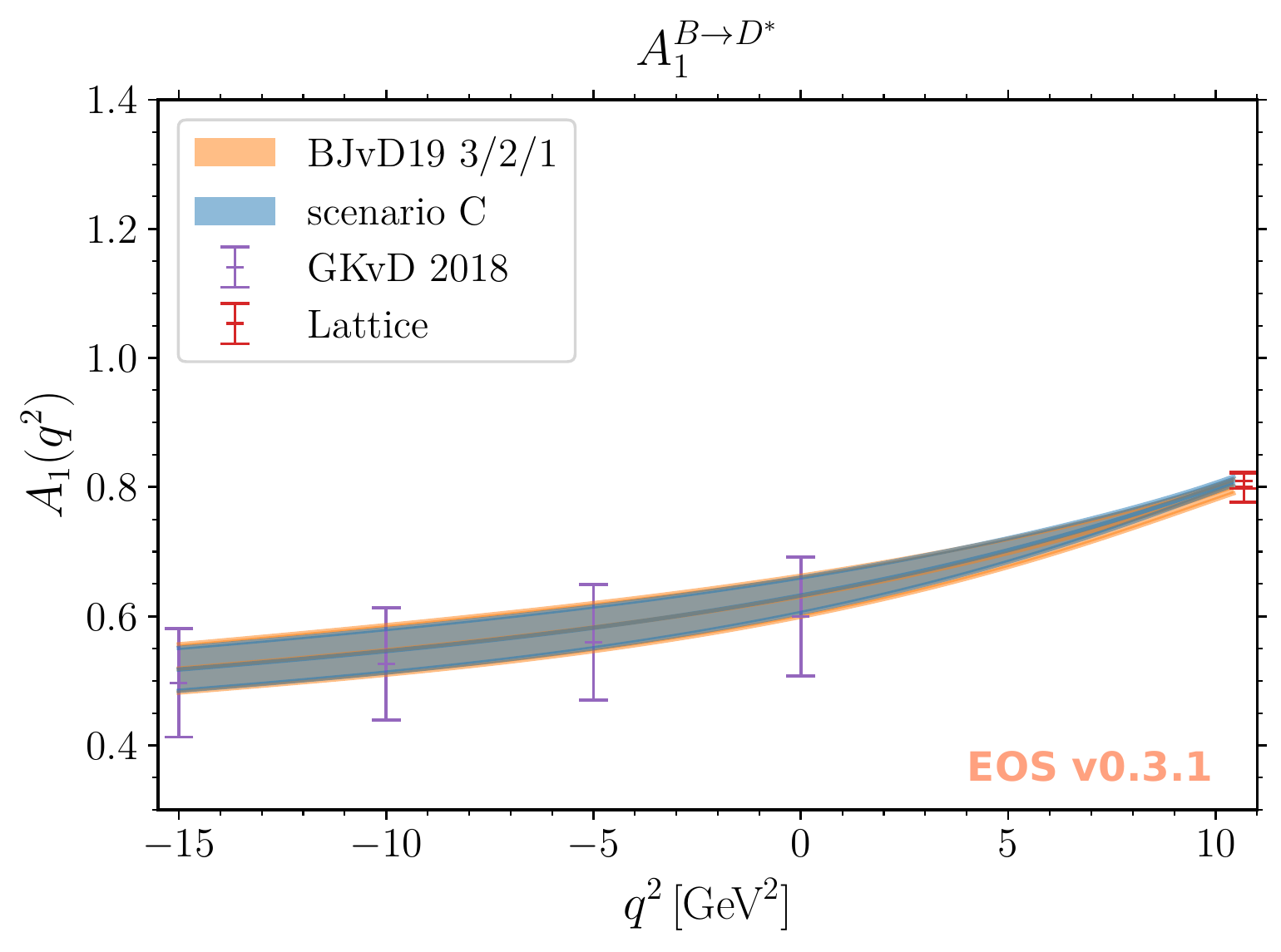}  &
        \includegraphics[width=.35\textwidth]{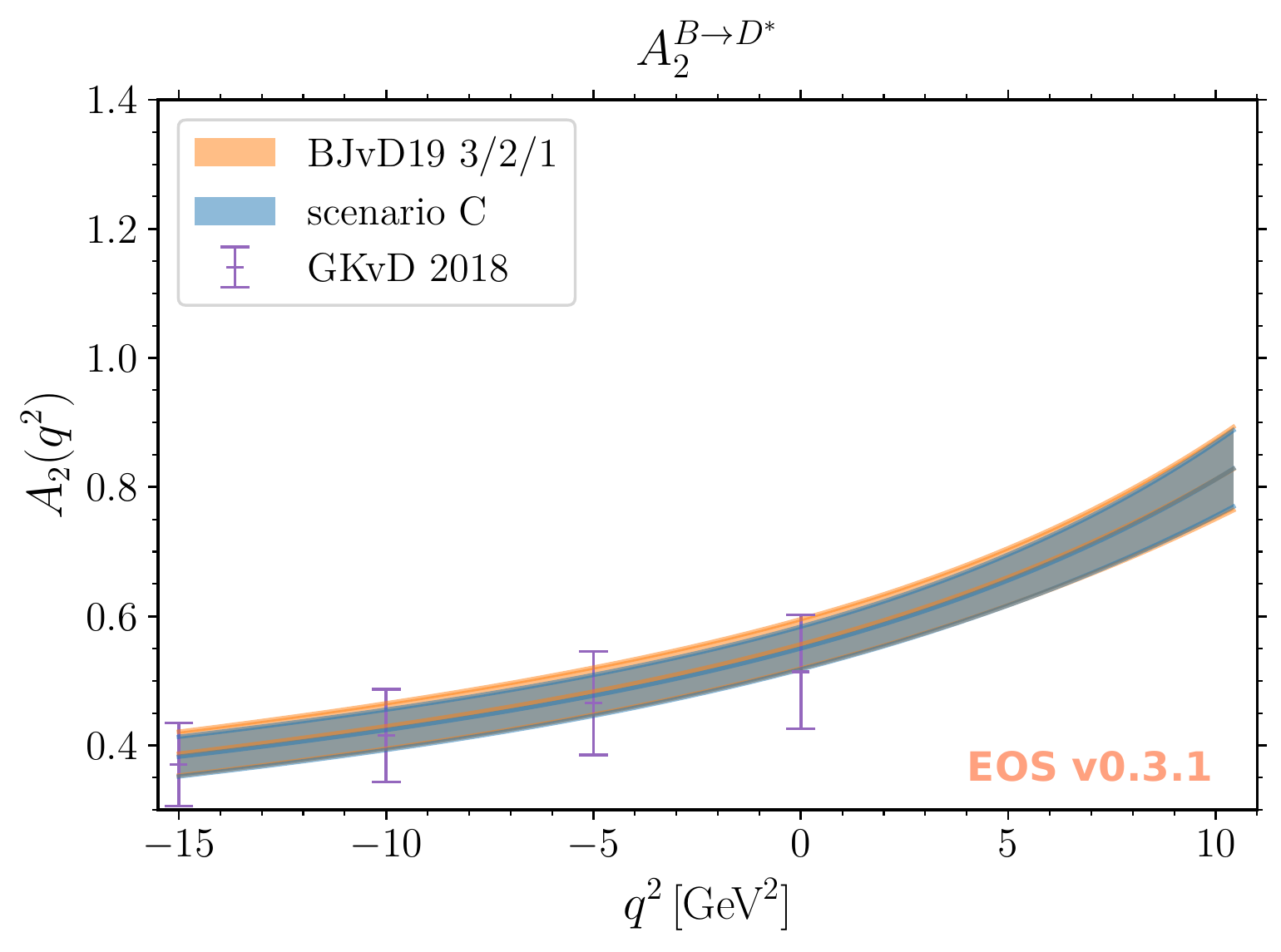}  \\[1.25em]
        \includegraphics[width=.35\textwidth]{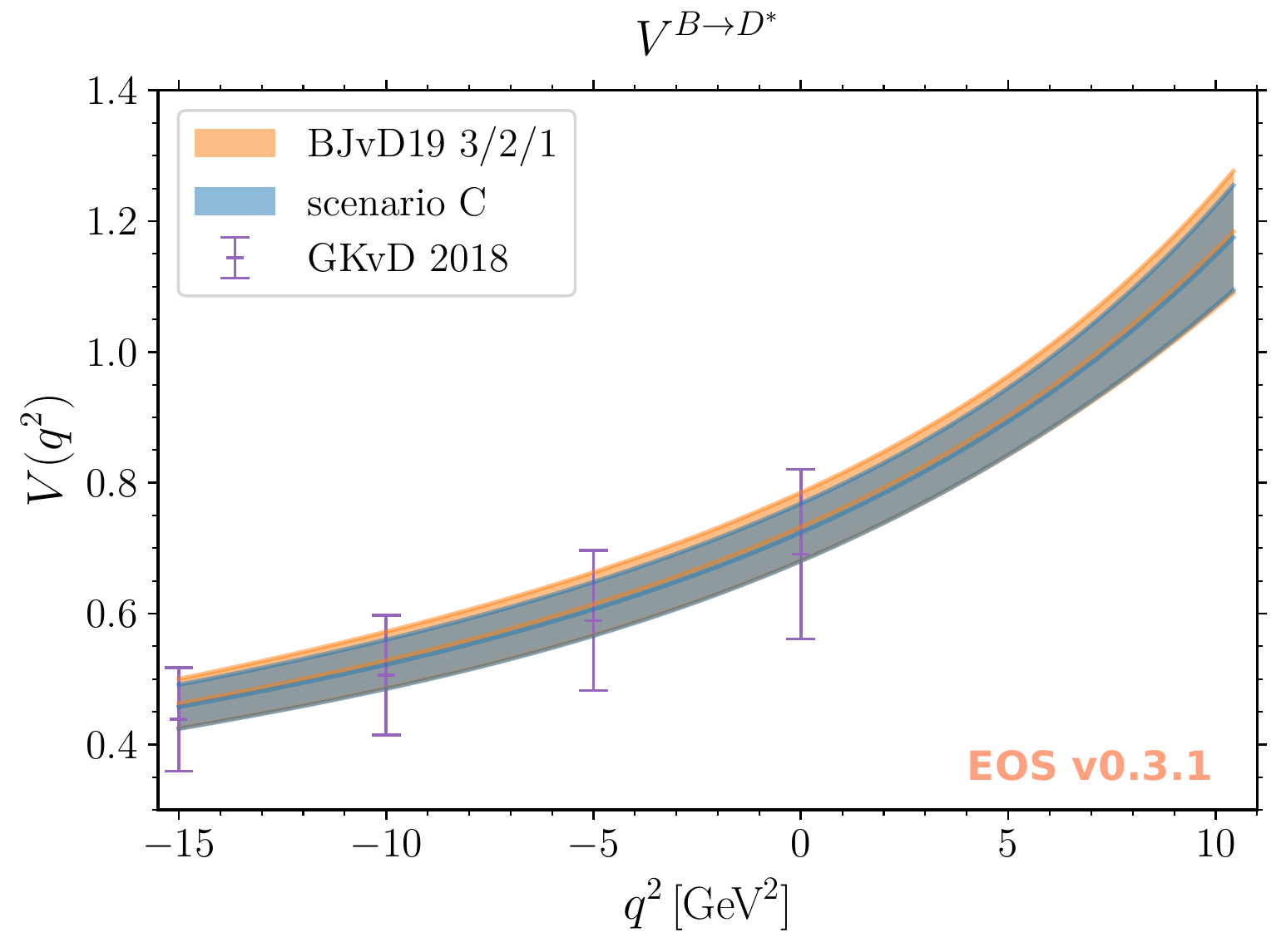}   &
        \includegraphics[width=.35\textwidth]{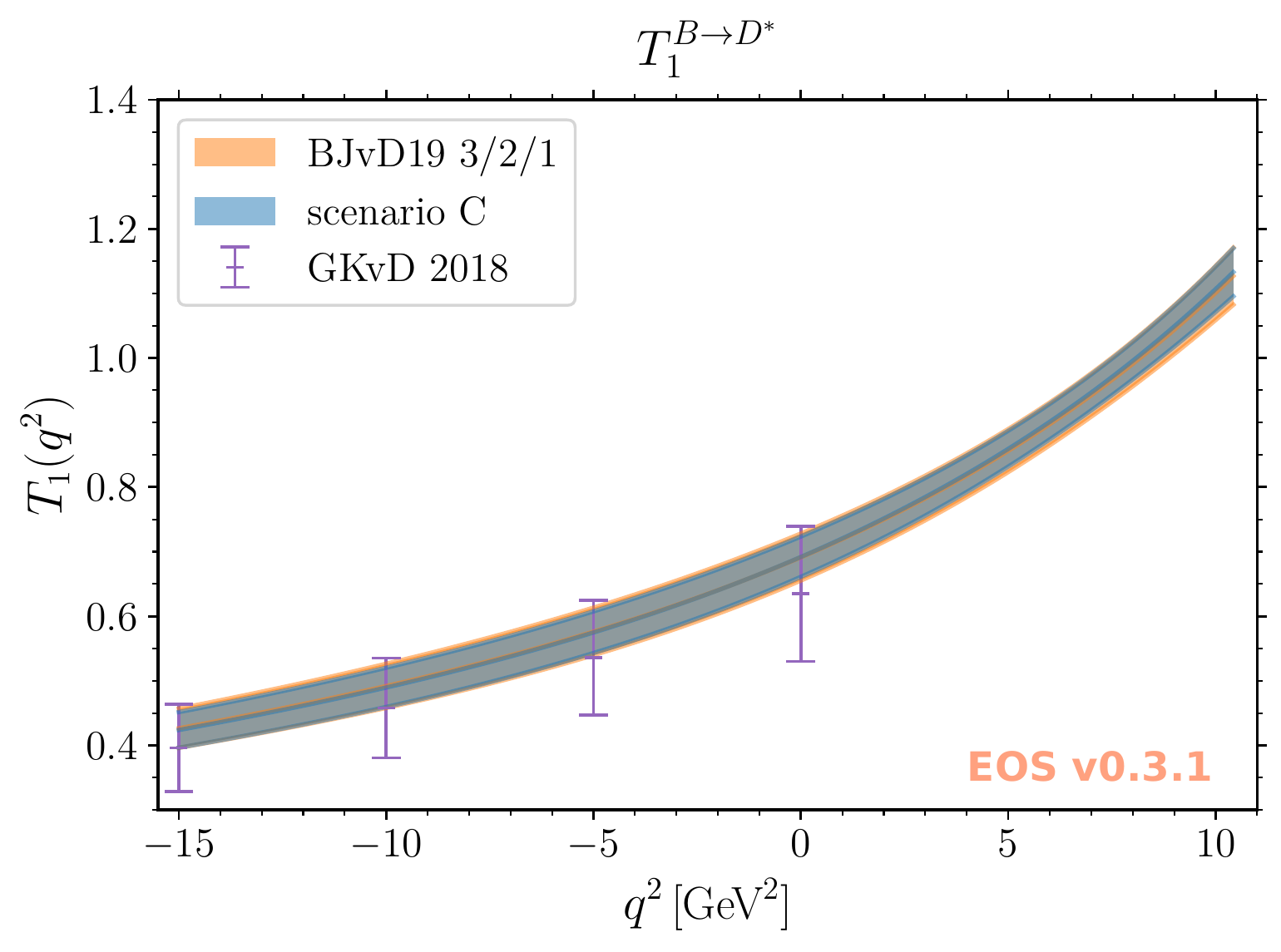}  &
        \includegraphics[width=.35\textwidth]{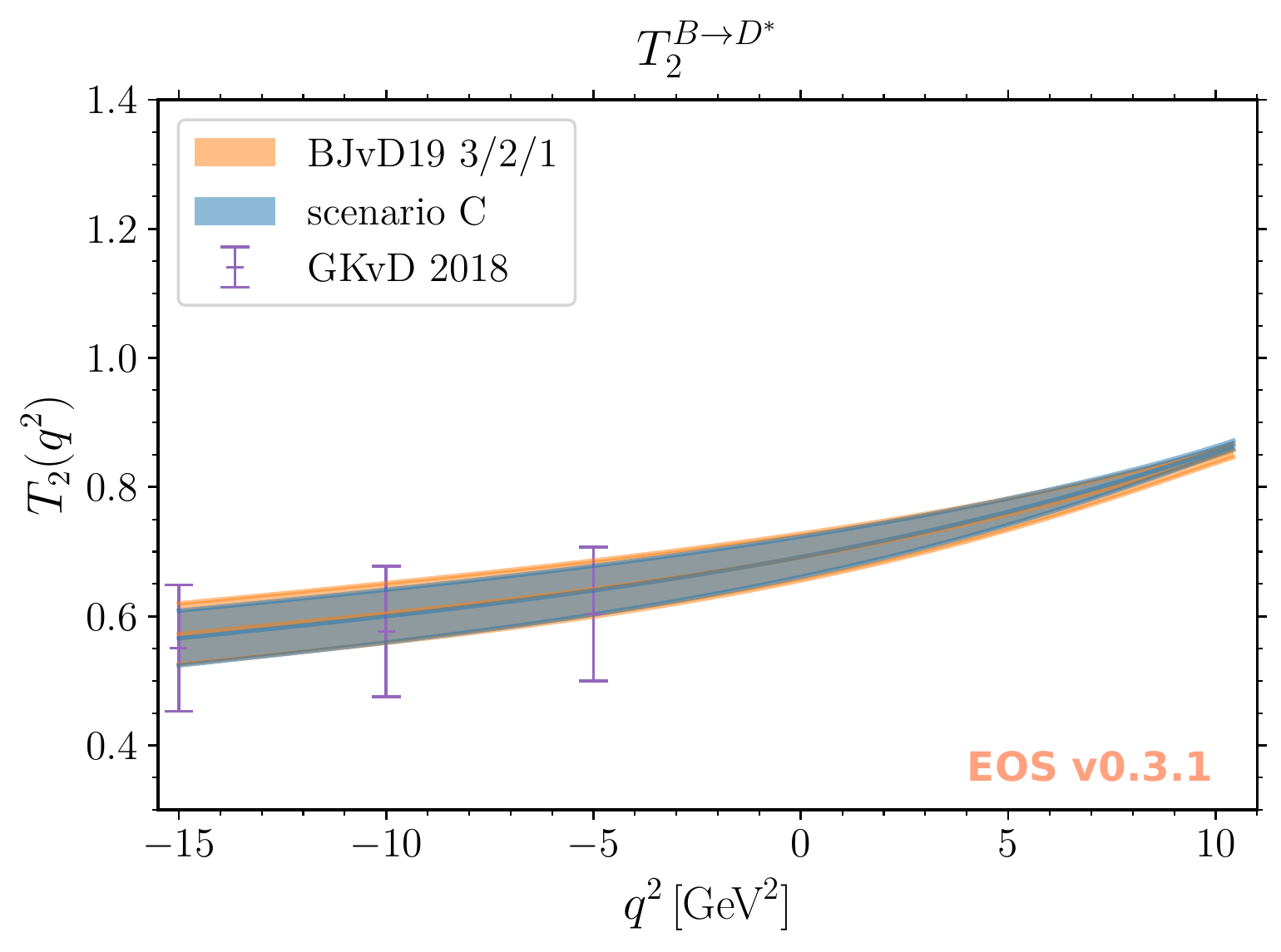}  \\[1.25em]
        \includegraphics[width=.35\textwidth]{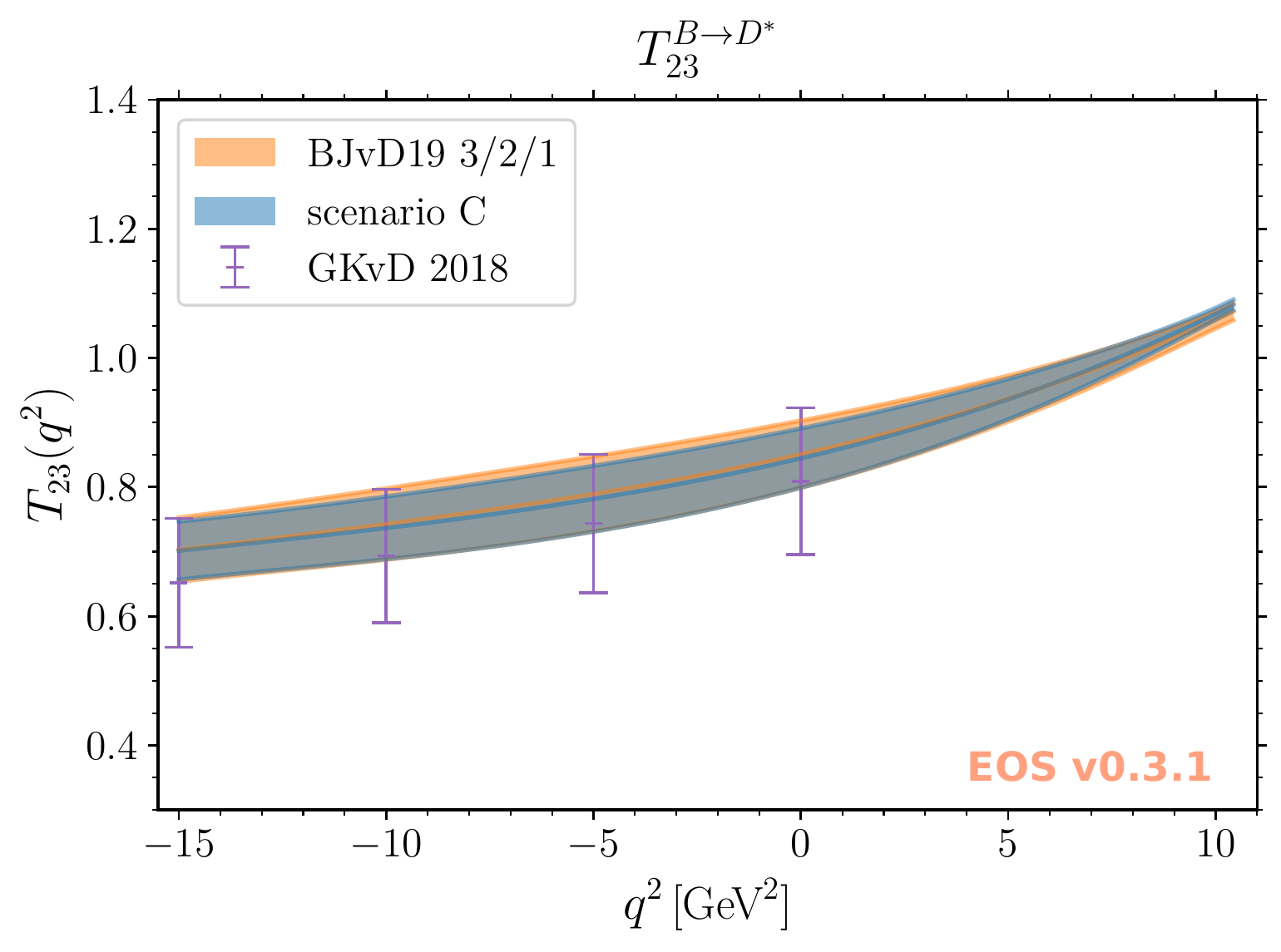}
    \end{tabular}
    \caption{
        The full set of $\bar{B}\to D^{(*)}$ form factors as a function of $q^2$ are used to showcase
        the agreement between the nominal results of ref.~\cite{Bordone:2019vic} (orange lines and areas)
        and the results when fully accounting for $\bar{B}_s^{(*)}\to D_s^{(*)}$ in the unitarity
        bounds (light blue lines and areas).
        For both sets of results we show the central values and $68\%$ probability envelopes from posterior-predictive
        distributions of the respective fits.
        The LCSR results taken from ref.~\cite{Gubernari:2018wyi} (purple points) and lattice constraints (red points) used in the fits are also shown in the plots.
    }
    \label{fig:B-to-DDstar-comparison}
\end{figure}

\textbf{Form factor predictions} With the posterior samples obtained from the fit to the nominal theory likelihood in scenario C, we produce
posterior predictive distributions for all $\bar{B}\to D^{(*)}$ form factors. In \reffig{B-to-DDstar-comparison},
the median curves and envelopes at $68\%$ probability are juxtaposed with those obtained using the $3/2/1$
model results in ref.~\cite{Bordone:2019vic}. We find very good agreement between the respective predictions, and obtain
slightly smaller uncertainties than in ref.~\cite{Bordone:2019vic}.
The largest change appears in the form factor $A_0$ close to zero recoil, with scenario C preferring slightly
larger values for this form factor than the previous analysis. The main reason for this is a strong saturation
of the unitarity bound in the $0^-$ channel due to our removing of the implicit assumption of $20\%$ $SU(3)_F$
breaking at the amplitude level.\\

\begin{table}[t!]
    \renewcommand{\arraystretch}{1.1}
    \begin{tabular}{l c cc cc cc cc}
        \toprule
        order
            & function $f$
            & \multicolumn{2}{c}{$f(1)$}
            & \multicolumn{2}{c}{$f'(1)$}
            & \multicolumn{2}{c}{$f''(1)$}
            & \multicolumn{2}{c}{$f'''(1)$}
            \\
        \midrule
        \multirow{2}{*}{$1/m_Q^0$}
            & $\xi$
            & $+1.00$ & ---
            & $-1.15$ & $[-1.30, -0.98]$
            & $+2.02$ & $[+1.64, +2.43]$
            & $-3.90$ & $[-4.90, -3.04]$
            \\
        \cmidrule{2-10}
        ~
            & $\xi^{(s)}$
            & $+1.00$ & ---
            & $-1.13$ & $[-1.38, -0.86]$
            & $+2.01$ & $[+1.46, +2.68]$
            & $-3.97$ & $[-5.54, -2.64]$
            \\
        \midrule
        \multirow{6}{*}{$1/m_Q^1$}
            & $\hat\chi_2$
            & $-0.07$ & $[-0.10, -0.03]$
            & $-0.02$ & $[-0.05, +0.02]$
            & $-0.01$ & $[-0.16, +0.18]$
            & ---     & ---
            \\
        ~
            & $\hat\chi_3$
            & $+0.00$ & ---
            & $+0.04$ & $[+0.01, +0.07]$
            & $-0.11$ & $[-0.17, -0.06]$
            & ---     & ---
            \\
        ~
            & $\hat\eta$
            & $+0.64$ & $[+0.50, +0.79]$
            & $+0.06$ & $[-0.16, +0.27]$
            & $-0.52$ & $[-1.08, +0.01]$
            & ---     & ---
            \\
        \cmidrule{2-10}
        ~
            & $\hat\chi_2^{(s)}$
            & $-0.07$ & $[-0.10, -0.03]$
            & $-0.00$ & $[-0.03, +0.04]$
            & $+0.15$ & $[-0.22, +0.57]$
            & ---     & ---
            \\
        ~
            & $\hat\chi_3^{(s)}$
            & $+0.00$ & ---
            & $+0.03$ & $[+0.01, +0.07]$
            & $-0.13$ & $[-0.23, -0.03]$
            & ---     & ---
            \\
        ~
            & $\hat\eta^{(s)}$
            & $+0.68$ & $[+0.53, +0.83]$
            & $-0.12$ & $[-0.37, +0.15]$
            & $-0.73$ & $[-1.75, +0.25]$
            & ---     & ---
            \\
        \midrule
        \multirow{6}{*}{$1/m_Q^2$}
            & $\hat\ell_1$
            & $+0.17$ & $[-0.02, +0.37]$
            & $-5.80$ & $[-11.6, -0.59]$
            & ---     & ---
            & ---     & ---
            \\
        ~
            & $\hat\ell_2$
            & $-1.60$ & $[-1.82, -1.37]$
            & $-3.73$ & $[-8.43, +0.76]$
            & ---     & ---
            & ---     & ---
            \\
        ~
            & $\hat\ell_3$
            & $-3.52$ & $[-9.41, +2.49]$
            & $+5.12$ & $[-0.04, +10.4]$
            & ---     & ---
            & ---     & ---
            \\
        ~
            & $\hat\ell_4$
            & $-2.33$ & $[-3.54, -1.14]$
            & $-0.72$ & $[-2.55, +1.05]$
            & ---     & ---
            & ---     & ---
            \\
        ~
            & $\hat\ell_5$
            & $+3.04$ & $[+1.00, +5.10]$
            & $+0.18$ & $[-2.01, +2.46]$
            & ---     & ---
            & ---     & ---
            \\
        ~
            & $\hat\ell_6$
            & $+2.33$ & $[-0.64, +5.40]$
            & $+0.70$ & $[-2.45, +3.96]$
            & ---     & ---
            & ---     & ---
            \\
        \bottomrule
    \end{tabular}
    \renewcommand{\arraystretch}{1.0}
    \caption{Best-fit point for the parameters of scenario C in a simultaneous fit to theory constraints and all
        available experimental measurements. Uncertainty ranges presented here are meant for illustrative purpose only,
        and should not be interpreted as standard deviations due to non-Gaussianity of the joint posterior.
    }
    \label{tab:BFP}
\end{table}

\begin{figure}[p]
    \begin{tabular}{ccc}
        \includegraphics[width=.35\textwidth]{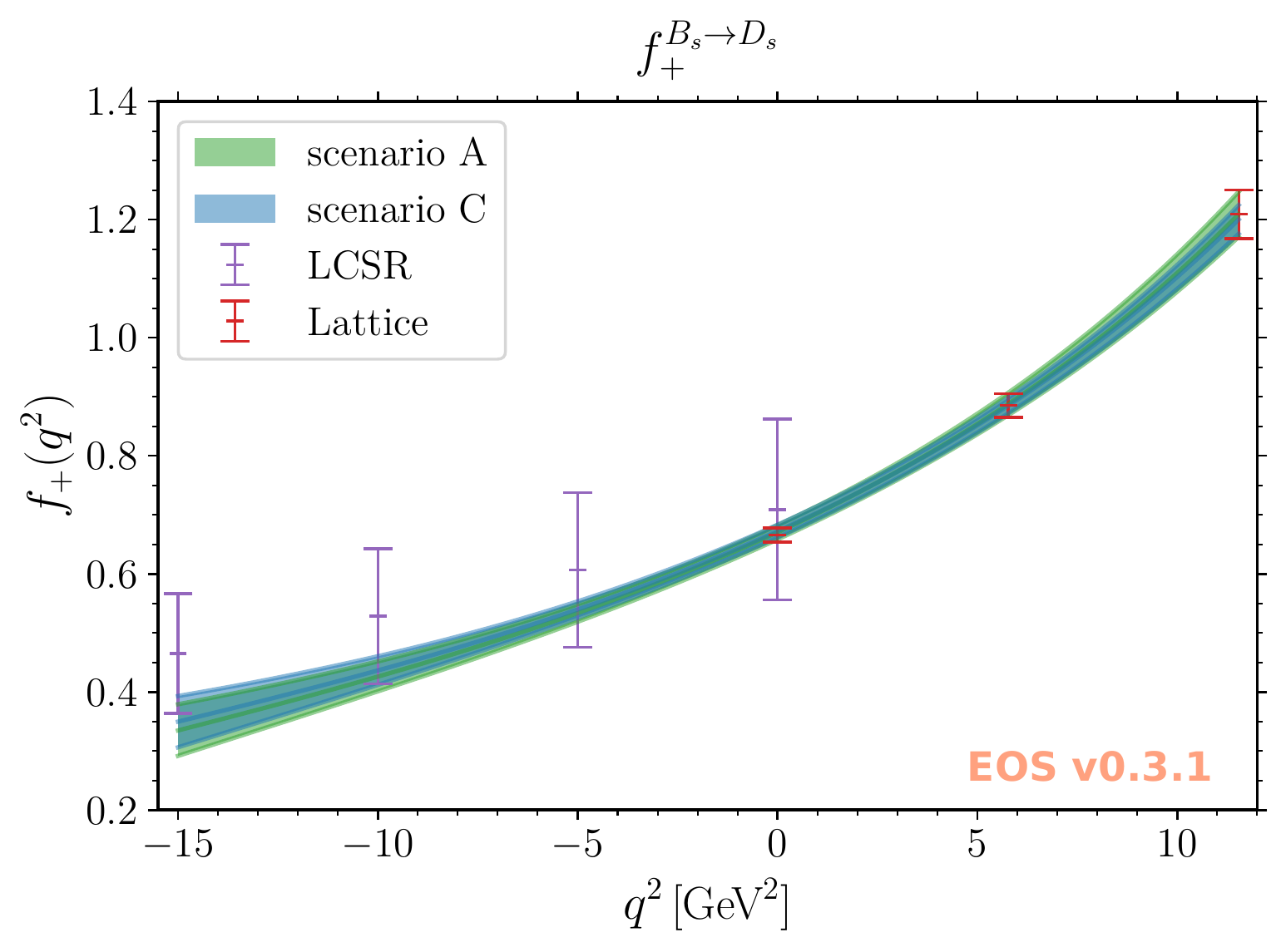}  &
        \includegraphics[width=.35\textwidth]{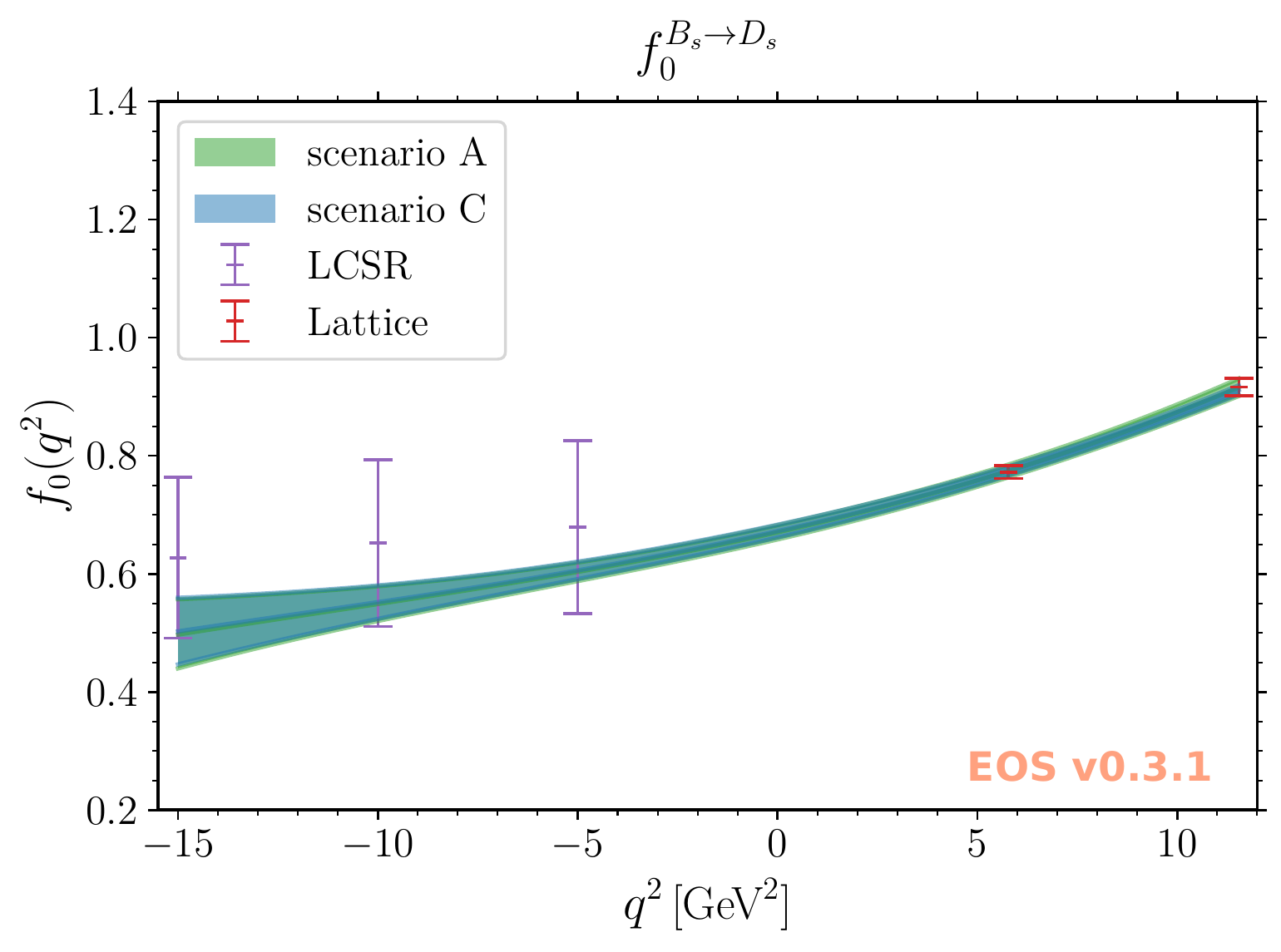}  &
        \includegraphics[width=.35\textwidth]{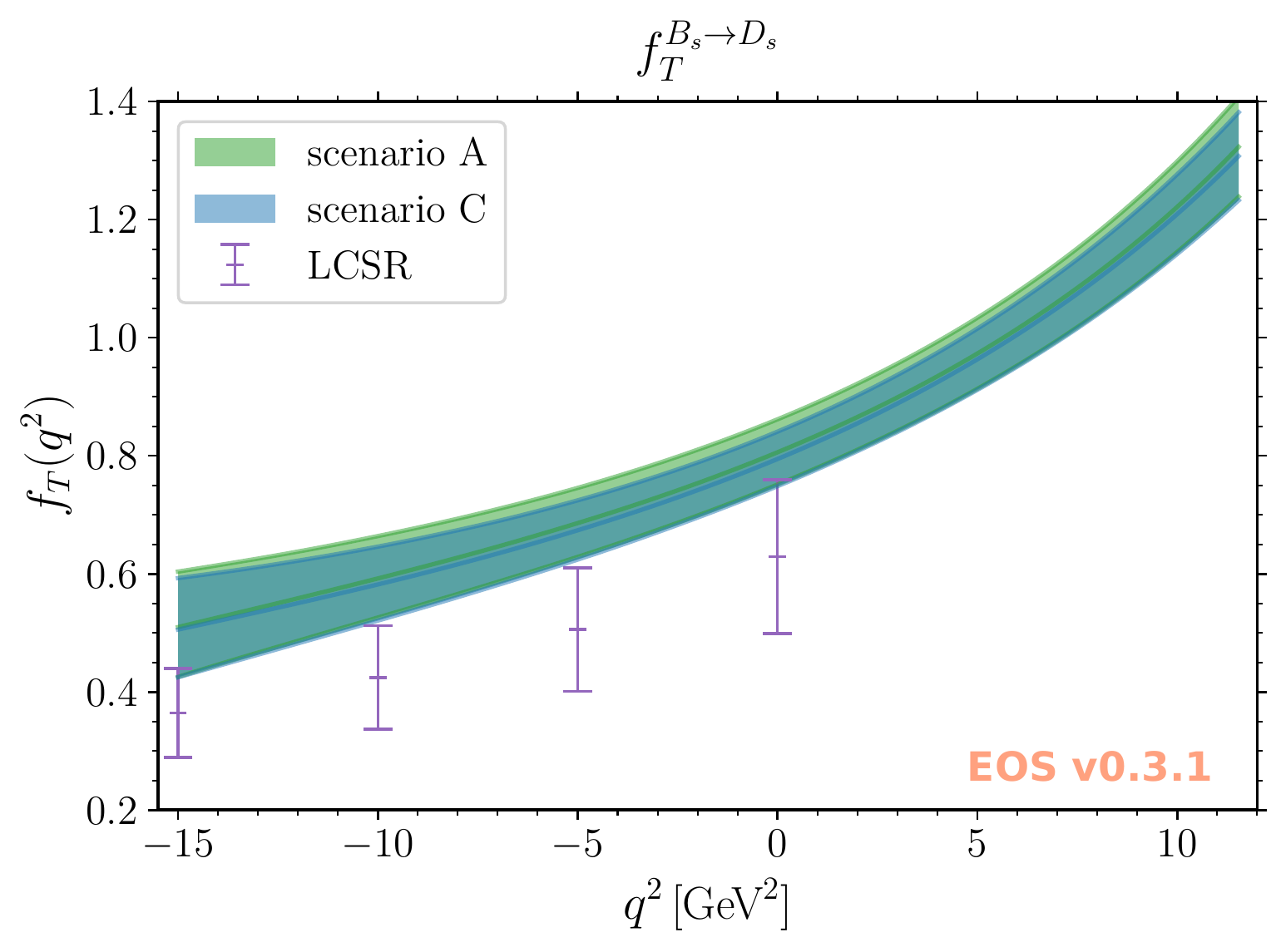}  \\[1.25em]
        \includegraphics[width=.35\textwidth]{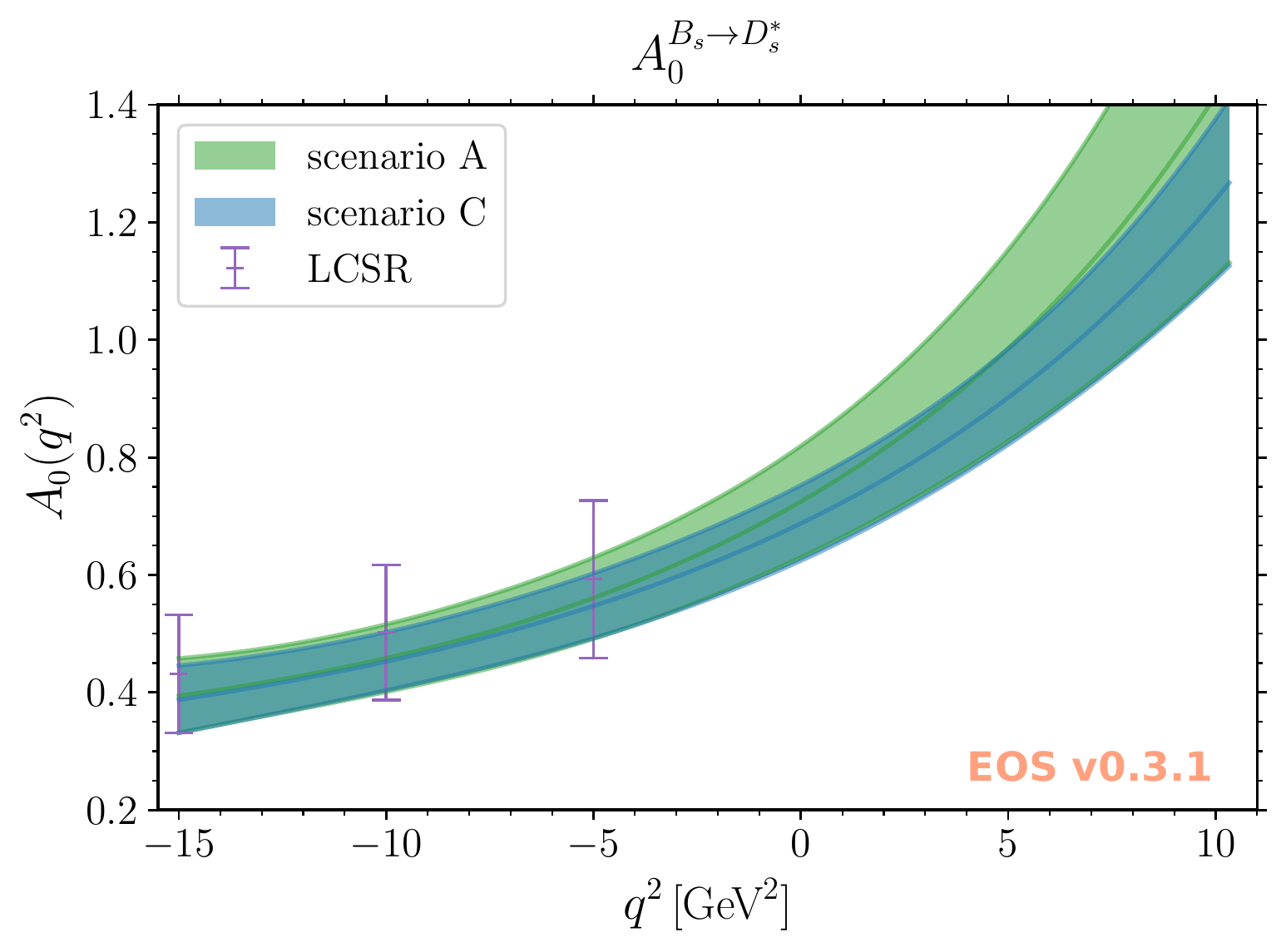}  &
        \includegraphics[width=.35\textwidth]{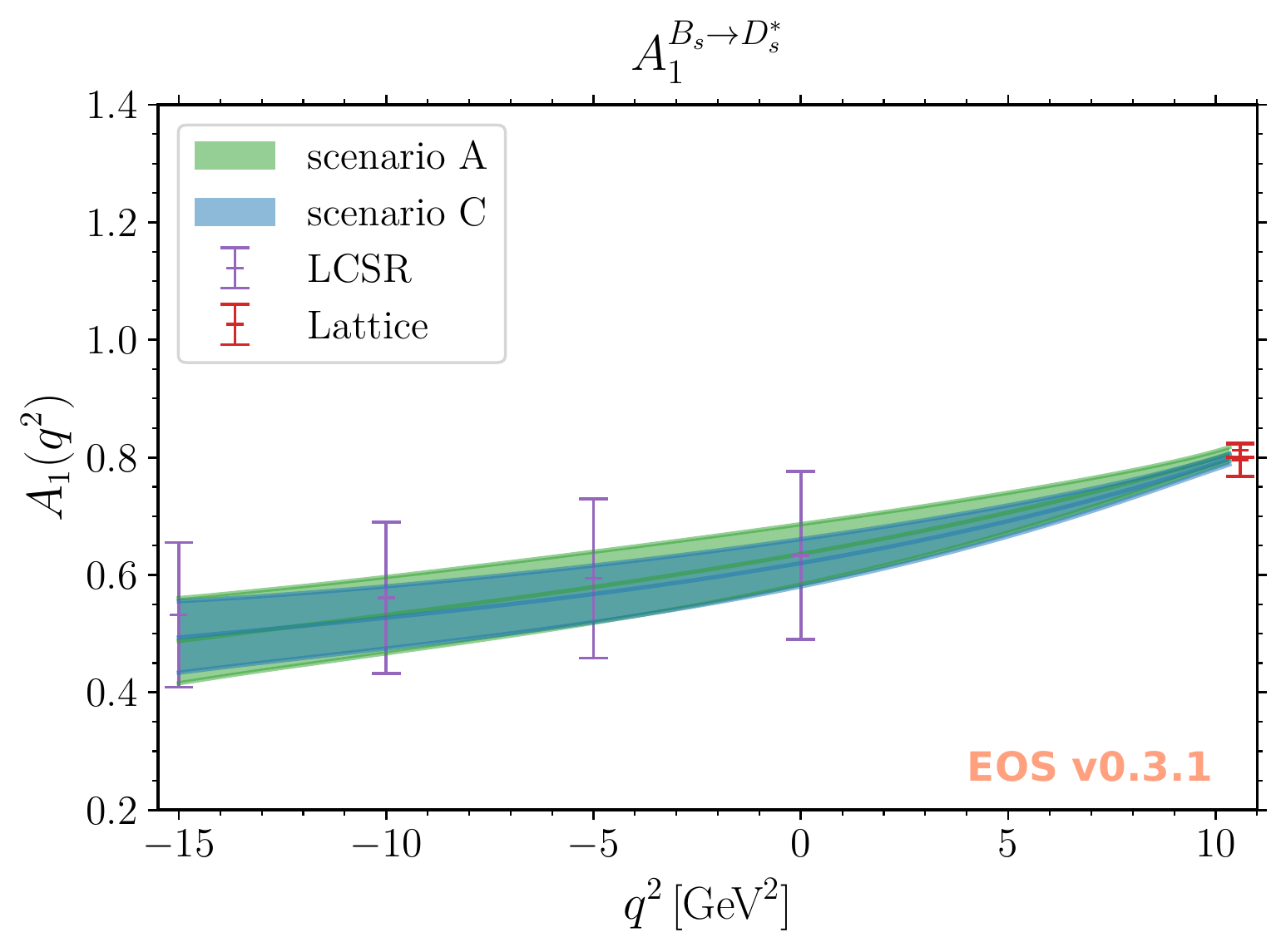}  &
        \includegraphics[width=.35\textwidth]{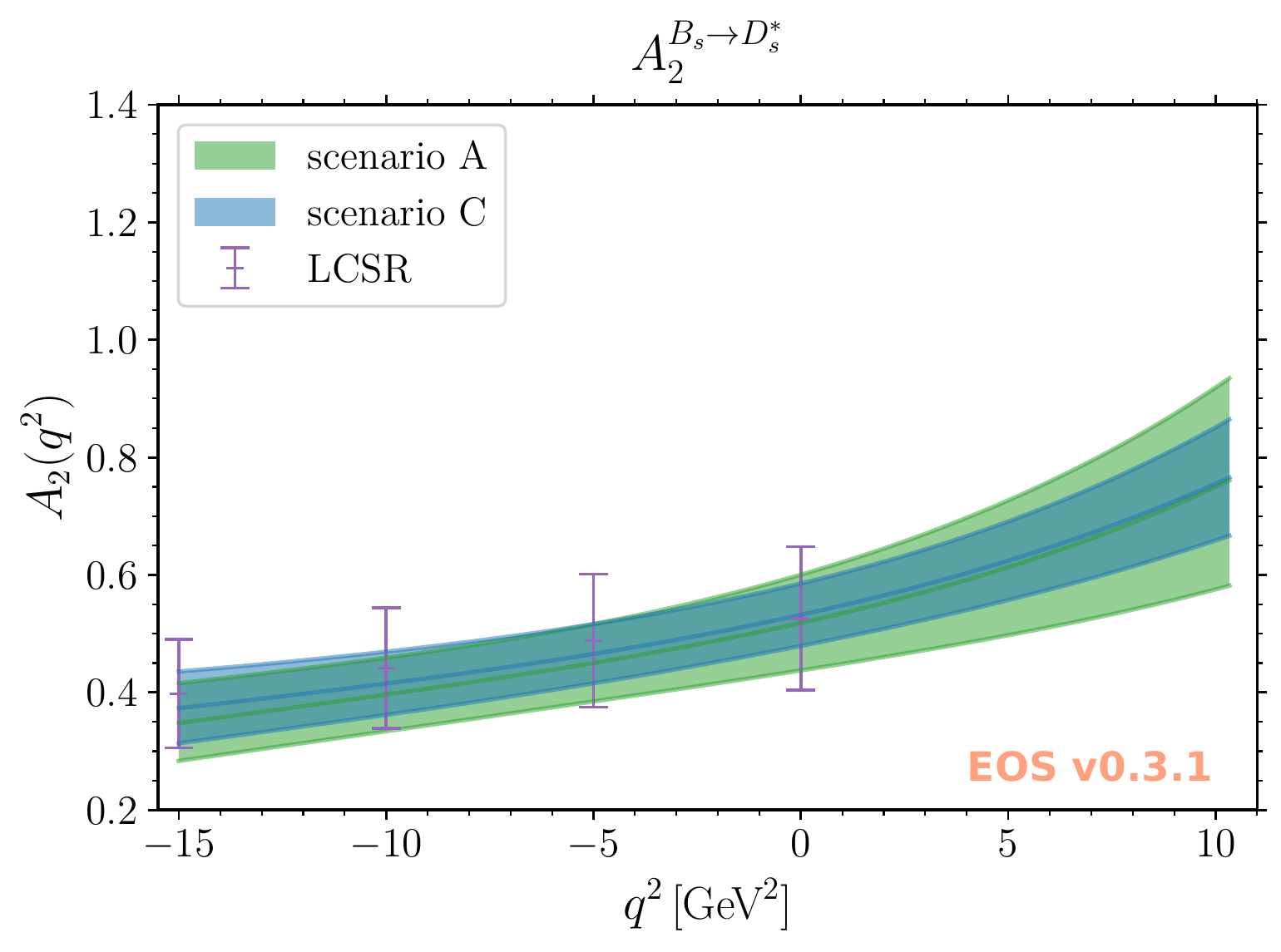}  \\[1.25em]
        \includegraphics[width=.35\textwidth]{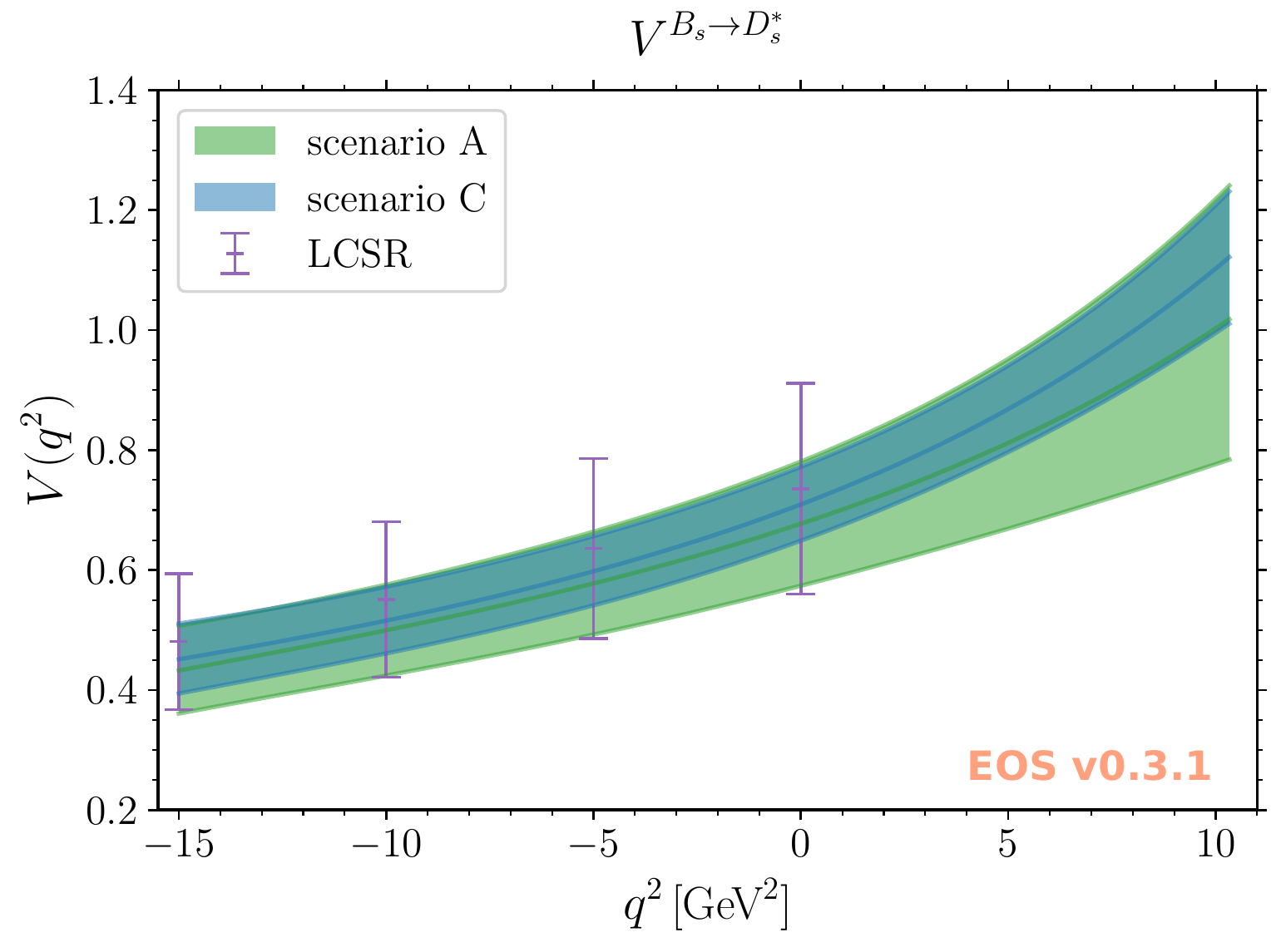}   &
        \includegraphics[width=.35\textwidth]{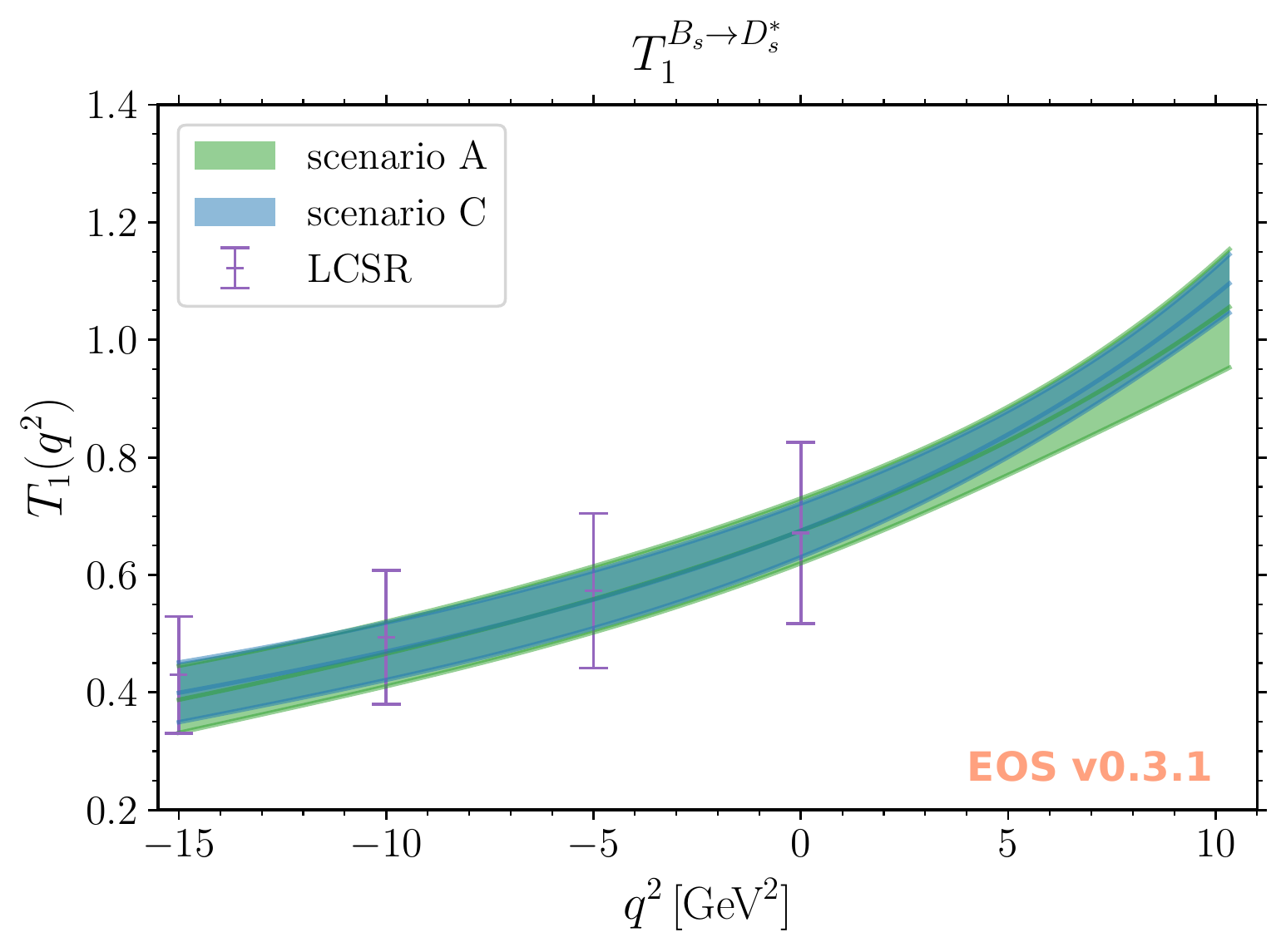}  &
        \includegraphics[width=.35\textwidth]{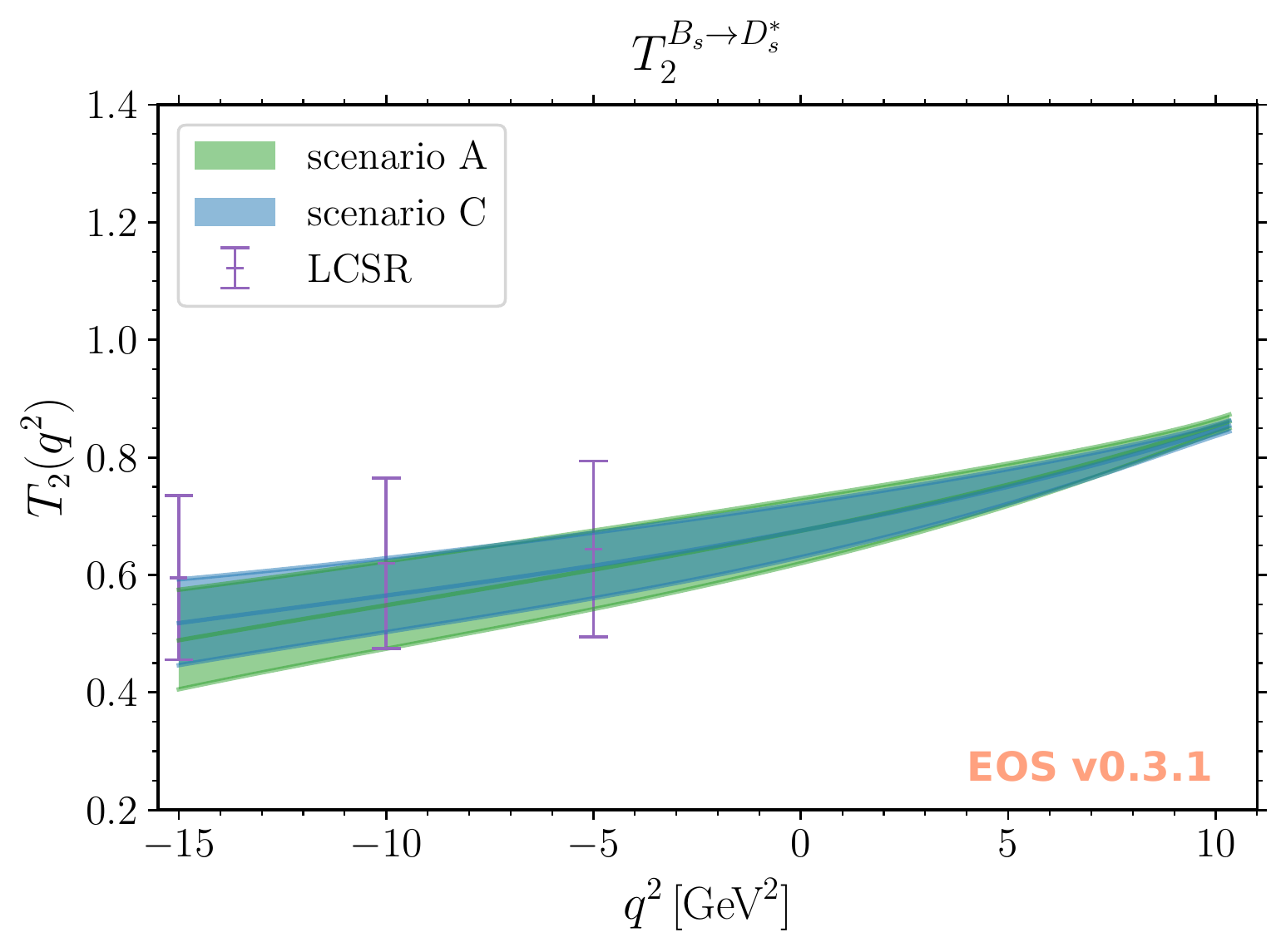}  \\[1.25em]
        \includegraphics[width=.35\textwidth]{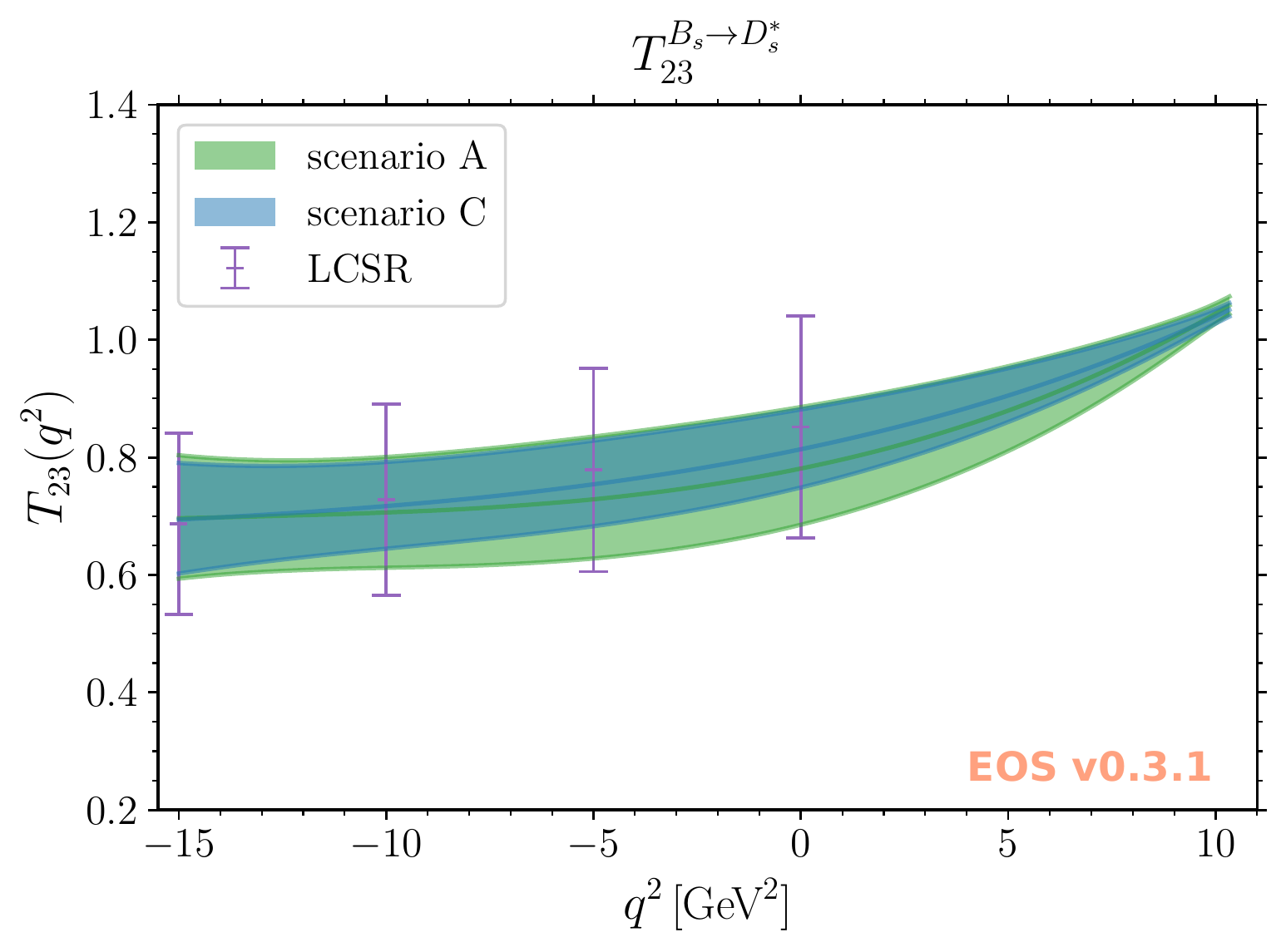}
    \end{tabular}
    \caption{
        Comparison between scenario A (green lines and areas) and scenario C (light blue lines and areas) of $\bar{B}_s\to D_s^{(*)}$ form factors as a function of $q^2$.
        For both sets of results we show the central values and $68\%$ probability envelopes from posterior-predictive distributions of the respective fits.
        The LCSR (purple points, see \refapp{LCSRs}) and lattice QCD constraints used in the fits (red points, see refs.~\cite{Bailey:2014tva,Na:2015kha,Harrison:2017fmw,Bailey:2014tva,Aoki:2019cca}) are also shown in the plots.
    }
    \label{fig:Bs-to-DsDsstar-comparison}
\end{figure}

We use the posterior samples from scenario A and scenario C to produce posterior predictive distributions
for all $\bar{B}_s\to D_s^{(*)}$ form factors. Their respective median curves and $68\%$ probability envelopes are
juxtaposed in \reffig{Bs-to-DsDsstar-comparison}. The fit using scenario C significantly reduces
the uncertainty of the $\bar{B}_s\to D_s^{(*)}$ form factors when compared to scenario A. This is an expected result,
since in scenario C the subsubleading IW are shared between $q=u,d$ and $q=s$ spectators, and the $SU(3)_F$ breaking is taken into account explicitly in the unitarity bounds instead of weakening them by a rough estimate for the breaking.\\

\textbf{BGL coefficients} The type of form factor parametrization put forward in ref.~\cite{Boyd:1995cf} allows for a straight-forward application of the unitarity
constraints and allows to put a strict upper limit on higher-order contributions. We produce posterior predictive distributions for the coefficients of this
parametrization, and find
their joint distribution to be accurately represented by a multivariate Gaussian distribution.
Within ancillary files attached to this preprint we provide the $17$ independent coefficients up to
order $z^2$ separately for the $\bar{B}$ and $\bar{B}_s$ decays. We exclude $a_0^{A_5}$, which can be obtained
as
\begin{align}
 \frac{a_0^{A_5}}{a_0^{A_1}}=2\sqrt{2}\frac{1-\sqrt{r}}{1+\sqrt{r}}\,,
\end{align}
where $r=m_{D_q^*}/m_{\bar{B}_q}$.
Both sets of coefficients are provided exclusively within scenario C, for the two fits to either the nominal theory likelihood,
or to the combined theory and experimental likelihood.\footnote{%
    Note that these values do not obey the equation of motion $f_0(q^2=0)=f_+(q^2=0)$ exactly.
    If desired, one of the BGL coefficients in this relation can be replaced by the remaining ones
    in order to enforce this identity. The correlation matrix for the joined set of $\bar B_q\to D_q^{(*)}$
    form factors becomes non-trivial due to additional relations between the BGL coefficient to the order
    in $1/m_{c,b}$ we are working in. This matrix can be obtained on request.
}
\\

\begin{figure}[t!]
    \begin{tabular}{ccc}
        \includegraphics[width=.35\textwidth]{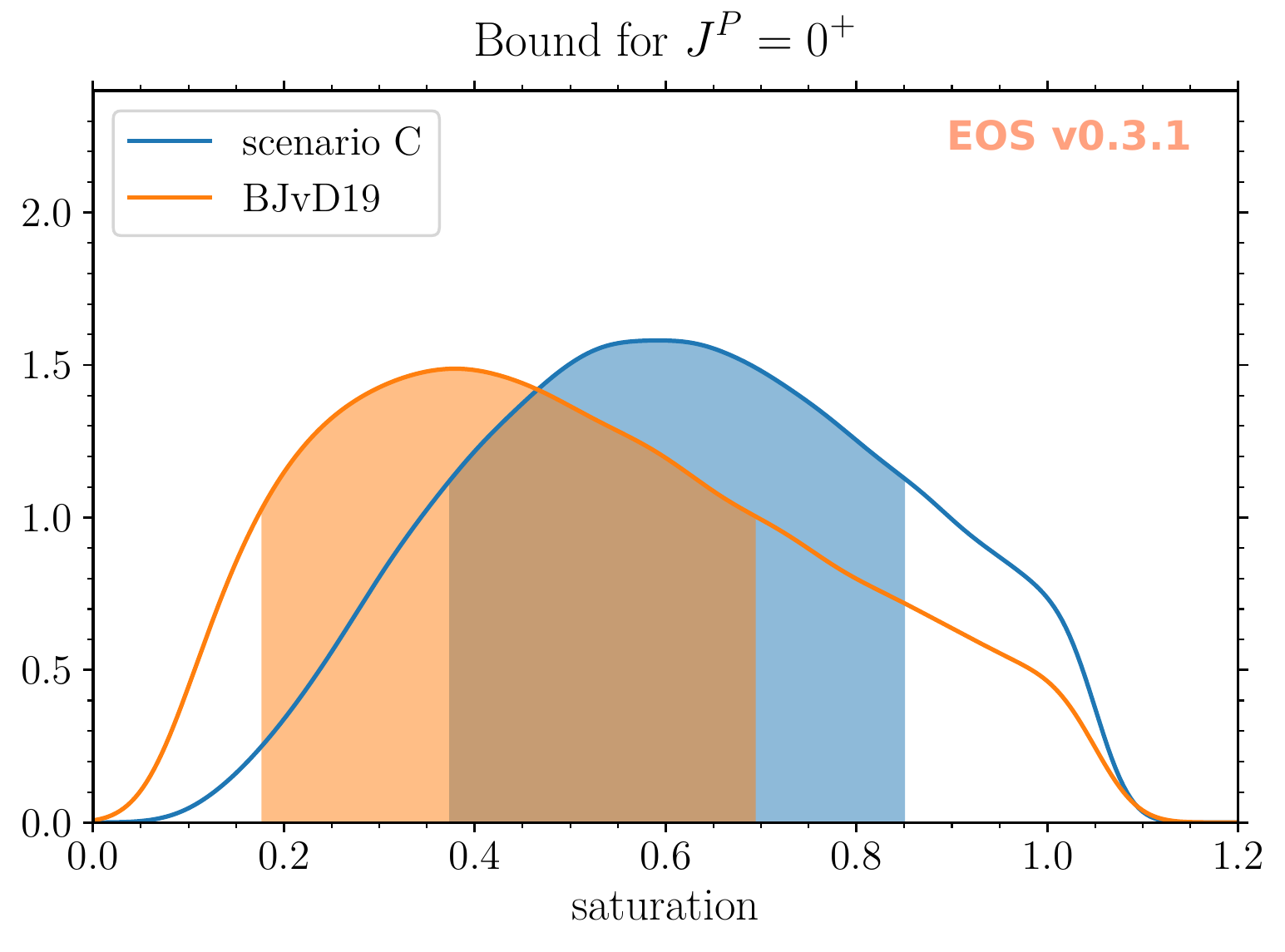}  &
        \includegraphics[width=.35\textwidth]{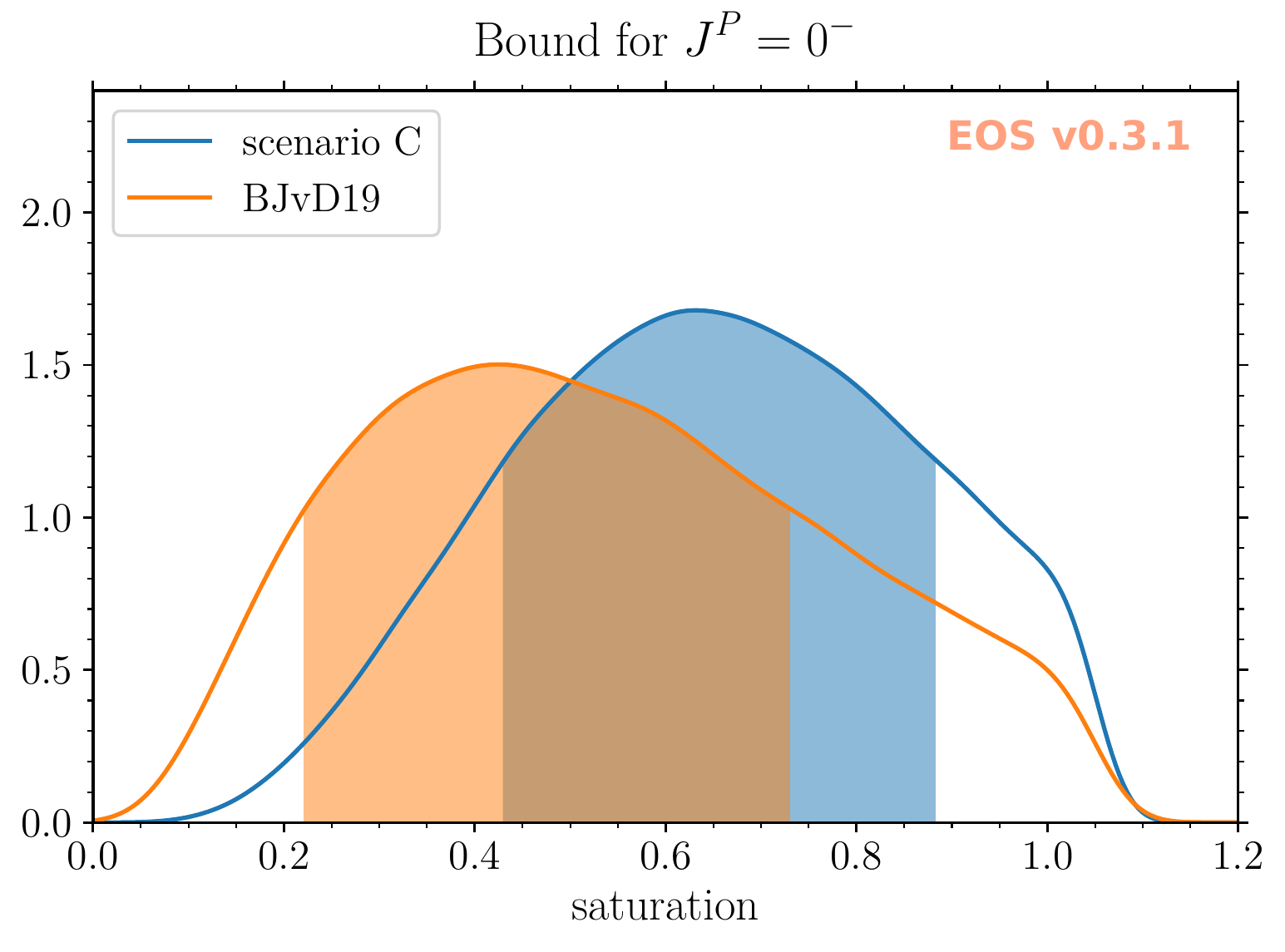}  \\[1.25em]
        \includegraphics[width=.35\textwidth]{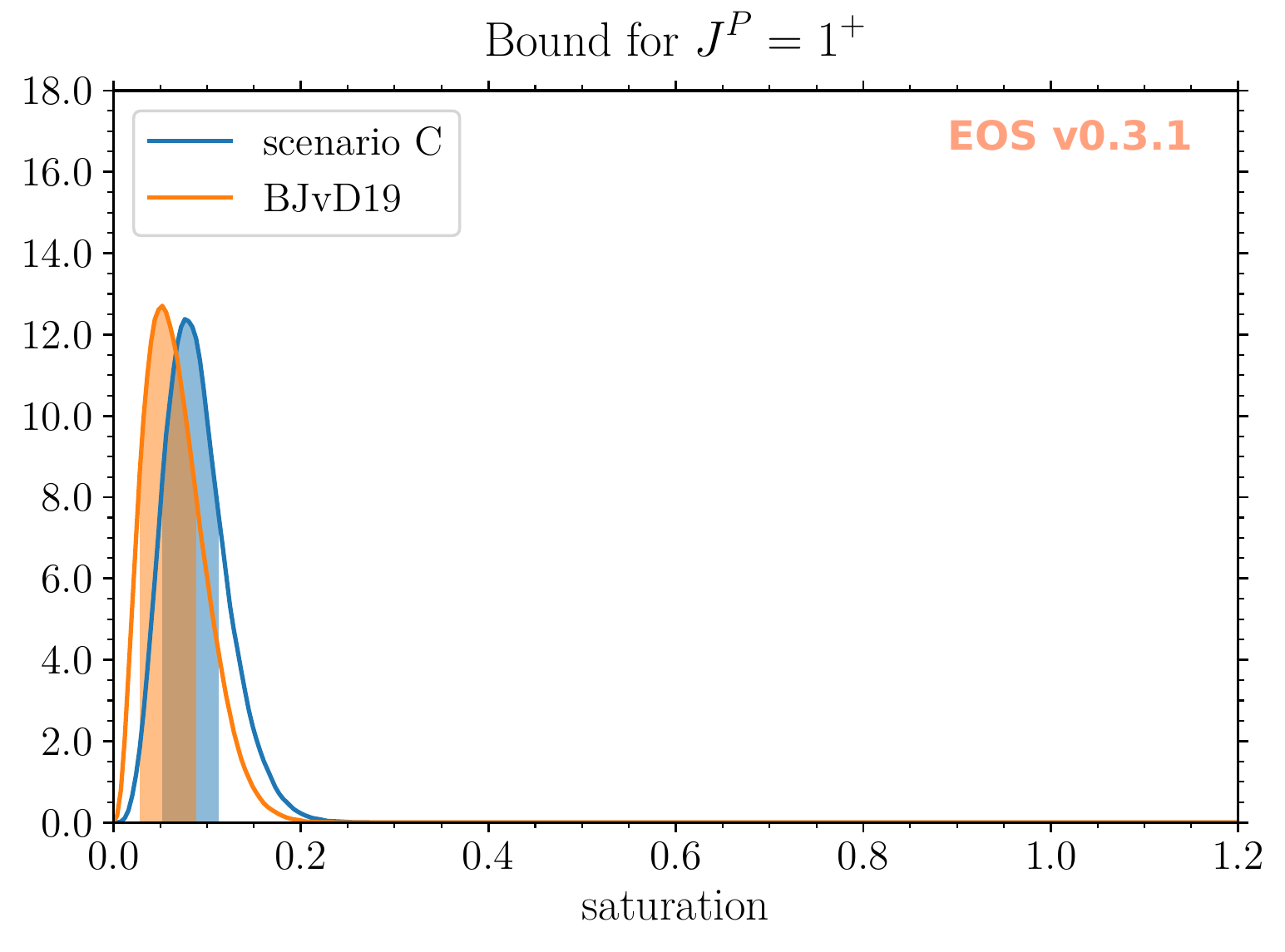}  &
        \includegraphics[width=.35\textwidth]{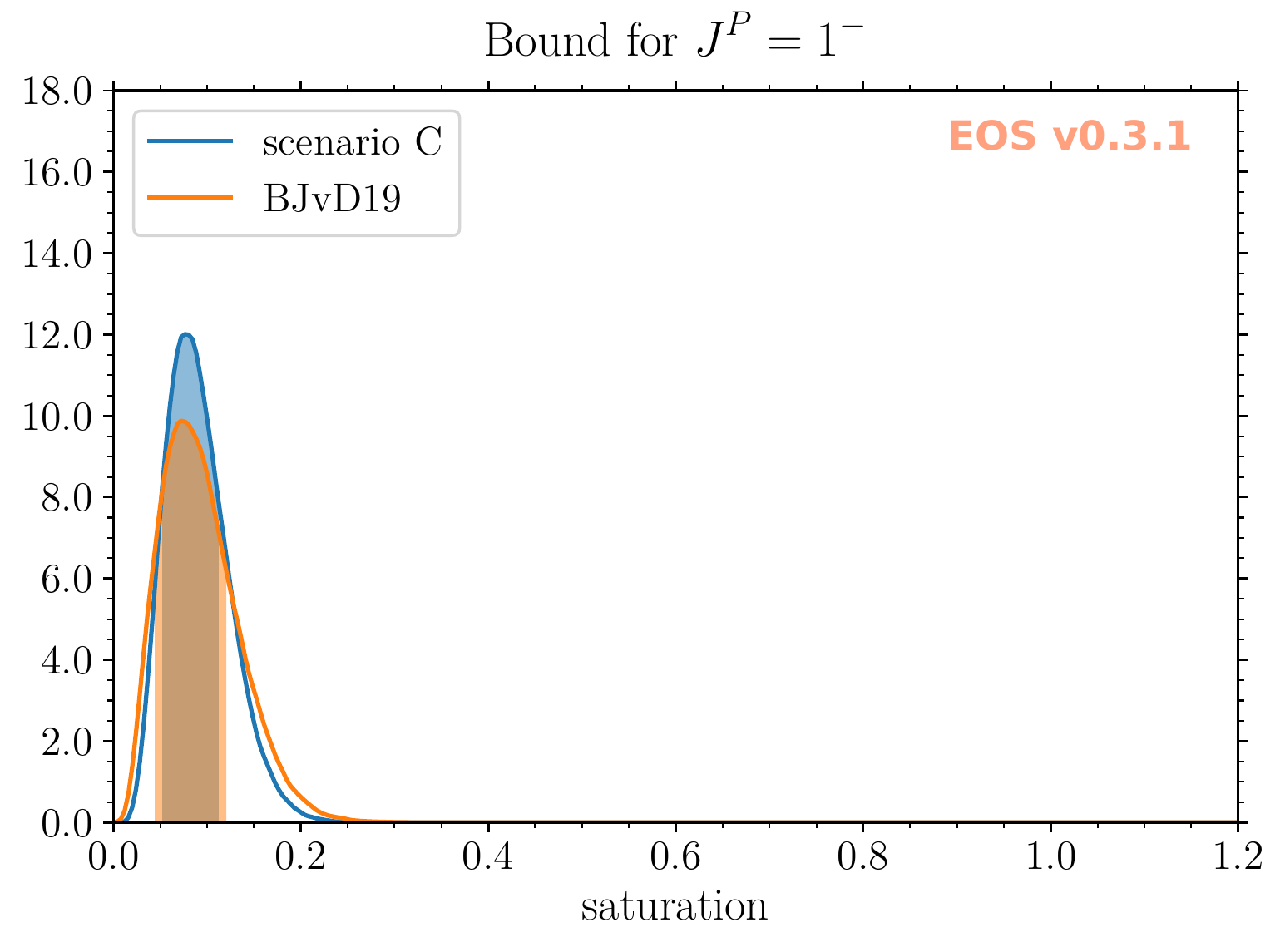}  \\
    \end{tabular}
    \caption{
        Posterior predictive distributions for the unitarity bounds in the channels $J^P=0^+$, $0^-$, $1^+$, and $1^-$. We compare the results from this work (blue lines and shaded areas) with the results obtained in ref.~\cite{Bordone:2019vic} (orange lines and shaded areas).
    }
    \label{fig:Bounds}
\end{figure}

\textbf{Effects on the unitarity bounds} The question that started this analysis is regarding the saturation of the unitarity bounds when forgoing the assumption of $SU(3)_F$ symmetry. We find the best-fit points of ref.~\cite{Bordone:2019vic} to fully saturate the bounds
of that analysis, as does the best-fit point of our scenario C in this analysis. This can be understood, since the bounds represent a non-linear
prior on the HQE parameter space, and the likelihoods exhibit their global minimum outside of the support of this prior. The bounds can therefore be saturated to $100\%$, which poses another question:
How likely is any given level of saturation of the bounds? This question can best be answered by computing the posterior predictive
distributions of the contributions to the unitarity bounds. These distributions yield the probability density of each bound
within the model description, given the available data. In \reffig{Bounds}, we juxtapose the results obtained for the
3/2/1 model in ref.~\cite{Bordone:2019vic} with our results in scenario C. We find that the mode, \emph{i.e.}, the most likely
level of saturation, in both the $0^+$ and the $0^-$ bounds increases from $\sim 0.4$ to $\sim 0.6$.
At the amplitude level this represents a relative increase of $22\%$. Including the results for $q=s$ spectators in the bounds
therefore increases the average saturation of the bounds, and yields further and significant restrictions on the HQE parameter space.
The prevailing assumption of \emph{reducing} the $q=s$ contributions to the unitarity bounds by $20\%$ should
therefore be abandoned for future analyses. We obtain the following median values and central $68\%$ intervals are for the four channels:
\begin{equation}
\begin{aligned}
    (J^P = 0^+)
        & ~         &
    \text{median}
        & = 0.62\,, &
    68\%\,\,\text{interval}
        & : [0.37, 0.85]\,, \\
    (J^P = 0^-)
        & ~         &
    \text{median}
        & = 0.65\,, &
    68\%\,\,\text{interval}
        & : [0.43, 0.88]\,, \\
    (J^P = 1^+)
        & ~         &
    \text{median}
        & = 0.08\,, &
    68\%\,\,\text{interval}
        & : [0.05, 0.11]\,, \\
    (J^P = 1^-)
        & ~         &
    \text{median}
        & = 0.09\,, &
    68\%\,\,\text{interval}
        & : [0.05, 0.11]\,. \\
\end{aligned}
\end{equation}
With this result, we illustrate that the unitarity bounds for the scalar and pseudoscalar currents
have now become an indispensable ingredient in fitting any data on the form factors within the HQE.
We note in passing that we find no significant shifts in the saturation of the bounds when using the
combined theoretical and experimental likelihood.\\

\textbf{Predictions for the lepton-flavour universality (LFU) ratios} With the posterior samples of the various fits in hand, we can proceed to produce posterior predictive distributions for
the LFU ratios.\\

Within the fit of scenario A to the likelihood comprised only of theory predictions of $\bar{B}_s\to D_s^{(*)}$
matrix elements, we obtain for the median values and central $68\%$ probability intervals:
\begin{equation}
\begin{aligned}
    R(D_s)
        & = 0.2979 \pm 0.0044\,,   &
    R(D_s^*)
        & = 0.246 \pm 0.010\,,
\end{aligned}
\end{equation}
with negligible correlation between both results.\\

Within the fit of scenario C to the nominal theory-only likelihood, we obtain the following
as the median values and central $68\%$ probability intervals of all four LFU
ratios:
\begin{equation}
\begin{aligned}
    R(D)
        & = 0.2989 \pm 0.0032\,,   &
    R(D_s)
        & = 0.2970 \pm 0.0034\,,   \\
    R(D^*)
        & = 0.2472 \pm 0.0050\,,   &
    R(D_s^*)
        & = 0.2450 \pm 0.0082\,.
\end{aligned}
\end{equation}
In this way, we obtain central values of the theory prediction of $R(D_s)$ and $R(D_s^*)$ which are lower than the results in scenario A by $0.7\%$.
At the same time, we reduce the uncertainty of these predictions by $25$--$30\%$. The predictions for $R(D)$ and $R(D^*)$ stay virtually the same when compared to the theory-only results
of ref.~\cite{Bordone:2019vic}, with shifts smaller than $0.4\%$ and only a small reduction in the uncertainty of $R(D^*)$. The correlation matrix for our theory predictions is:
\begin{equation}
\newcommand{\pp}{\phantom{+}}
\begin{pmatrix}
    \pp 1.0000 & \pp 0.1257 & \pp 0.1294 & -   0.0205 \\
    \pp 0.1257 & \pp 1.0000 & \pp 0.0031 & \pp 0.3826 \\
    \pp 0.1294 & \pp 0.0031 & \pp 1.0000 & \pp 0.0016 \\
    -   0.0205 & \pp 0.3826 & \pp 0.0016 & \pp 1.0000         
\end{pmatrix}\,,
\end{equation}
in the order $R(D)$, $R(D^*)$, $R(D_s)$, and $R(D_s^*)$. Generally, the strongest correlations arise between the states with equal spin,
since identical combinations of IW functions enter. However, since potential correlations in the lattice QCD data for $\bar{B}\to D$ and $\bar{B}_s\to D_s$ are unknown,
the resulting correlations are smaller between these modes. The correlation between $R(D)$ and $R(D^*)$ results mainly from the LCSR results;
the corresponding one for $\bar{B}_s$ decays is much smaller and therefore so is the final correlation between $R(D_s)$ and $R(D_s^*)$.\\

When using the combined theoretical and experimental likelihood, we obtain
in the same way as above:
\begin{equation}
\begin{aligned}
    R(D)
        & = 0.2981 \pm 0.0029\,,   &
    R(D_s)
        & = 0.2971 \pm 0.0034\,,   \\
    R(D^*)
        & = 0.2504 \pm 0.0026\,,   &
    R(D_s^*)
        & = 0.2472 \pm 0.0077\,.
\end{aligned}
\end{equation}
The correlation matrix now reads:
\begin{equation}
\newcommand{\pp}{\phantom{+}}
\begin{pmatrix}
   \pp 1.0000  & \pp 0.0855  &\pp 0.1293  &    -0.0504 \\
   \pp 0.0855  & \pp 1.0000  &   -0.0132  & \pp 0.1768 \\
   \pp 0.1293  &    -0.0132  &\pp 1.0000  & \pp 0.0036 \\
      -0.0504  & \pp 0.1768  &\pp 0.0036  & \pp 1.0000              
\end{pmatrix}\,.
\end{equation}

\textbf{Polarizations in $\boldsymbol{\bar{B}_{q}\to D_{q}^{(*)}\tau^- \bar{\nu}}$} We produce posterior predictive distributions for the $\tau$ polarization $P_\tau^{D_{q}^{(*)}}$ in $\bar{B}_{q}\to D_{q}^{(*)}\tau^- \bar{\nu}$ decays and the longitudinal polarization fraction $F_L$ in $\bar{B}_{q}\to D_{q}^*\tau^- \bar{\nu}$ decays. In scenario C, using only theory constraints, we obtain
\begin{equation}
\begin{aligned}
    P_\tau^{D}     & = 0.3212 \pm 0.0029\,, &
    -P_\tau^{D^*}  & = 0.484  \pm 0.017 \,, & \\
                   &                        &
    F_L            & = 0.473  \pm 0.011 \,, & \\
    P_\tau^{D_s}   & = 0.3226 \pm 0.0096\,, &
    -P_\tau^{D_s^*}& = 0.477  \pm 0.025 \,, & \\
                 &                    &
    F^s_L          &=0.478 \pm 0.018 \,.
\end{aligned}
\end{equation}
Once we include the experimental PDFs in scenario C, we obtain
\begin{equation}
\begin{aligned}
    P_\tau^{D}     & = 0.3209\pm0.0029  \,, &
    -P_\tau^{D^*}  & = 0.492 \pm0.013   \,, & \\
                   &                    &
    F_L            & = 0.467 \pm 0.009  \,, & \\
    P_\tau^{D_s}   & = 0.3233\pm 0.0095 \,, &
    -P_\tau^{D_s^*}& = 0.486 \pm 0.023  \,, & \\
                 &                    &
    F^s_L          & = 0.471 \pm 0.016 \,.
\end{aligned}
\end{equation}

\textbf{Angular observables} Using the theory only fit within scenario C we predict the angular observables $J_i$ that
arise in the four-fold differential decay rate of $\bar{B}_q\to D_q^*\lbrace \mu^-, \tau^-\rbrace\bar\nu$ decays, see \emph{e.g.}\ ref.~\cite{Duraisamy:2014sna}.
Our results are presented using the same convention as in ref.~\cite{Feldmann:2015xsa}.
The central values with uncertainties are listed in \reftab{Js} while the correlation matrices are given as ancillary files attached to the arXiv preprint of this article.
While there are presently no measurements of the full set of these observables, an analysis strategy was recently suggested that allows to extract them
without model bias at the Belle II and LHCb experiments~\cite{Hill:2019zja}.\\

\textbf{Impact on $\boldsymbol{|V_{cb}|}$} We find reasonable agreement between the values of $|V_{cb}|$ extracted in
ref.~\cite{Bordone:2019vic} and in our analysis. When extracting from $\bar{B}\to D\ell^-\bar\nu$, $|V_{cb}|$ remains entirely
stable. When extracting $|V_{cb}|$ from $\bar{B}\to D^*\ell^-\bar\nu$. we observe a small downward shift in the simultaneous analysis
to the theory-only likelihood, which is almost entirely compensated when fitting to the combined likelihood.
We find that the compatibility with the inclusive determination worsens slightly to $1.8\sigma$ from the previous result of $1.2\sigma$.
We average the exclusive and inclusive determinations, and obtain
\begin{equation*}
    |V_{cb}| = (41.1 \pm 0.5) \cdot 10^{-3}\,,
\end{equation*}
which has smaller uncertainties than any of the previous determinations.

\begin{table}[t]
    \renewcommand{\arraystretch}{1.2}
    \begin{tabular}{l c c c c}
    \toprule
    ~                                                        &
        \multicolumn{4}{c}{scenarios}                        \\
    model                                                    &
        $3/2/1$                                              &
        $3/2/1$                                              &
        scenario C                                           &
        scenario C                                          \\
    exp.~likelihood                                          &
        ---                                                  &
        all exp.                                             &
        ---                                                  &
        all exp.                                             \\
    \midrule
    $\mathcal{B}(\bar{B}^0\to D^+\{e^-,\mu^-\}\bar\nu)/|V_{cb}|^2$   &
        $13.48 \pm 0.37$                                     &
        $13.56 \pm 0.35$                                     &
        $13.48 \pm 0.34$                                     &
        $13.54 \pm 0.32$                                     \\
    $\mathcal{B}(\bar{B}^0\to D^{*+}\{e^-,\mu^-\}\bar\nu)/|V_{cb}|^2$ &
        $33.16 \pm 2.15$                                     &
        $32.00 \pm 1.03$                                     &
        $33.87 \pm 1.82$                                     &
        $32.69 \pm 0.76$                                     \\
    correlation                                              &
        $0.14$                                               &
        $0.10$                                               &
        $0.13$                                               &
        $0.09$                                               \\
    \midrule
    $|V_{cb}|\times 10^3$ from $\bar{B}\to D\{e^-,\mu^-\}\bar\nu$   &
        $40.7 \pm 1.2$                                       &
        $40.6 \pm 1.1$                                       &
        $40.7 \pm 1.1$                                       &
        $40.7 \pm 1.1$                                           \\
    $|V_{cb}|\times 10^3$ from $\bar{B}\to D^*\{e^-,\mu^-\}\bar\nu$&
        $39.3 \pm 1.7$                                       &
        $40.0 \pm 1.1$                                       &
        $38.8 \pm 1.4$                                       &
        $39.5 \pm 0.9$                                       \\
    $|V_{cb}|\times 10^3$ combined incl. corr.               &
        $40.2 \pm 1.0$                                       &
        $40.3 \pm 0.8$                                       &
        $40.0 \pm 0.9$                                       &
        $40.0 \pm 0.7$                                       \\
    \midrule
    $\mathcal{B}(\bar{B}_s^0\to D_s^+\{e^-,\mu^-\}\bar\nu)/|V_{cb}|^2$   &
        ---                                                  &
        ---                                                  &
        $14.00 \pm 0.40$                                     &
        $13.99 \pm 0.40$                                     \\
    $\mathcal{B}(\bar{B}_s^0\to D_s^{*+}\{e^-,\mu^-\}\bar\nu)/|V_{cb}|^2$ &
        ---                                                  &
        ---                                                  &
        $33.04 \pm 2.88$                                     &
        $32.06 \pm 2.54$                                     \\
    correlation                                              &
        ---                                                  &
        ---                                                  &
        $-0.07$                                              &
        $-0.10$                                              \\
    \bottomrule
    \end{tabular}
    \renewcommand{\arraystretch}{1}
    \caption{%
        Branching ratios and $|V_{cb}|$ predictions from~\cite{Bordone:2019vic} and scenario C in this work, with and without experimental data.
    }
    \label{tab:BRs+Vcb}
\end{table}

\begin{table}[t]
    \renewcommand{\arraystretch}{1.2}
    \begin{tabular}{c r r r r}
    \toprule
    \diagbox{observable}{transition}  & 
    $\bar{B}\to D^*\mu^-\bar\nu$        &
    $\bar{B}\to D^*\tau^-\bar\nu$       &
    $\bar{B}_s\to D_s^*\mu^-\bar\nu$    &
    $\bar{B}_s\to D_s^*\tau^-\bar\nu$   \\   
    \midrule
    $J_1^s$                      &
    $0.257\pm 0.007$             &%J_1^s
    $0.279\pm 0.006$             &%J_1^s    
    $0.255\pm 0.012$             &%J_1^s    
    $0.277\pm 0.009$             \\%J_1^s     
    \midrule  
    $J_2^c$                      &
    $-0.399\pm 0.008$            &%J_2^c    
    $-0.128\pm 0.001$            &%J_2^c    
    $-0.402\pm 0.015$            &%J_2^c    
    $-0.127\pm 0.002$            \\%J_2^c     
    \midrule  
    $J_2^s$                      &
    $0.085\pm 0.002$             &%J_2^s    
    $0.047\pm 0.001$             &%J_2^s    
    $0.085\pm 0.004$             &%J_2^s    
    $0.047\pm 0.002$             \\%J_2^s    
    \midrule   
    $J_3$                        &
    $-0.133\pm 0.004$            &%J_3    
    $-0.082\pm 0.002$            &%J_3    
    $-0.135\pm 0.006$            &%J_3    
    $-0.082\pm 0.003$            \\%J_3    
    \midrule   
    $J_4$                        &
    $-0.230\pm 0.001$            &%J_4    
    $-0.105\pm 0.001$            &%J_4    
    $-0.231\pm 0.003$            &%J_4    
    $-0.105\pm 0.002$            \\%J_4    
    \midrule  
    $J_5$                        &
    $0.167\pm 0.008$             &%J_5   
    $0.207\pm 0.005$             &%J_5    
    $0.161\pm 0.012$             &%J_5
    $0.204\pm 0.007$             \\%J_5     
    \midrule  
    $J_6^c$                      &
    $0.011\pm 0.001$             &%J_6^c    
    $0.277\pm 0.015$             &%J_6^c    
    $0.011\pm 0.001$             &%J_6^c    
    $0.282\pm 0.023$             \\%J_6^c   
    \midrule    
    $J_6^s$                      & 
    $-0.203\pm 0.012$            &%J_6^s    
    $-0.163\pm 0.012$            &%J_6^s    
    $-0.194\pm 0.016$            &%J_6^s    
    $-0.155\pm 0.015$            \\%J_6^s 
    \bottomrule
    \end{tabular}
    \renewcommand{\arraystretch}{1}
    \caption{%
        Central values and uncertainties for the non-redundant and non-vanishing angular observables in $\bar{B}_q\to D_q^*\lbrace \mu^-, \tau^-\rbrace\bar\nu$
        in the Standard Model.
    }
    \label{tab:Js}
\end{table}

\section{Summary and outlook}
\label{sec:summary}

We present the first simultaneous analysis of the form factors in $\bar{B}_q \to D_q^{(*)}\ell^-\bar\nu$ decays with $q=u,d,s$
beyond the $SU(3)_F$ symmetry limit.
In addition to all available lattice QCD data our analysis makes use of two new sets of results, which have been produced for this work.
First, we include new light-cone sum rule (LCSR) results
for all $\bar{B}_s \to D_s^{(*)}\ell^-\bar\nu$ form factors except $f_T^{(s)}$, which are obtained close to and beyond the point of maximum recoil.
Second, we revisit the existing QCD sum rules for the subleading Isgur-Wise (IW) functions. We obtain a consistent set of predictions for both
light and strange spectator quarks. Our results for the light spectator are consistent with previous numerical results in the literature~\cite{Bernlochner:2017jka}.
A clear benefit of our simultaneous analysis is that we are no longer forced to make assumption about the $q=s$ form factors in the unitarity bounds.
\\

We consider three different fit scenarios, all of which fit our various datasets well. Scenario A is used only when fitting exclusively the form factors
for $q=s$ spectator quarks, and shows compatibility with $SU(3)_F$ symmetry in a first test. 
Scenarios B and C are used in simultaneous
fits to both $q=u,d$ and $q=s$ data. Scenario C is more constrained, since it assumes a combined power counting in which $1/m_c$ power corrections
in the HQE are of similar size as the $SU(3)_F$ breaking. Through a Bayesian model comparison, we find that 
scenario C is the most efficient in describing the available data, and we therefore only use this scenario to obtain all nominal results of
our analysis. Using 34 parameters, scenario C can predict a total of 20 form factors in the entire semileptonic phase space.
We make this information available through ancillary files, both for the parametrization in terms of parameters of the IW functions as well
as those in the BGL parametrization.
Our results include furthermore precise SM predictions for the branching ratios, lepton-flavour universality ratios, the complete non-redundant and non-vanishing
set of angular observables in $\bar{B}_q\to D_q^*\lbrace \mu^-, \tau^-\rbrace\bar\nu$ decays, and the tau polarizations.
We find good agreement between our results and the
results of ref.~\cite{Bordone:2019vic}, indicated by sub-percent shifts in the predictions of $q^2$ integrated observables. The precision of our predictions for observables in semileptonic $\bar{B}_s\to D_s^{(*)}$ decays
has now reached a similar level as the prediction for observables in semileptonic $\bar{B}_{u,d}\to D_{u,d}^{(*)}$ decays.\\

To obtain a better understanding of the structure of the unitarity bounds, we analyze posterior predictive distributions of the saturation
of the bounds. We find that our simultaneous analysis increases the median saturation compared to the previous analysis by $\sim 22\%$
at the amplitude level, which is of the same order as the naive reduction based on simple dimensional estimates of the $SU(3)_F$ breaking
used in previous applications of the unitarity bounds. This reflects the observation that the form factors are perfectly compatible with $SU(3)_F$ symmetry at the present level of precision. Combining the $q=u,d$ and $q=s$ likelihoods in a simultaneous fit shows clear benefits: first, the unitarity bounds
yield stronger constraints on the parameters space due to the larger degree of saturation. Second, the parametric uncertainties for
all the IW functions decrease, with the largest effects in the parameters of the subsubleading IW functions. The consequence of
both effects is a significant increase in the precision of the theory predictions of all observables considered in this work.
Moreover, our analysis will be able to serve as an important cross check of the upcoming lattice QCD results for $\bar{B}_q\to D_q^*$ form factors, which can subsequently be included in order to be used in the analysis of future measurements in these decays.

\textbf{Note added:} In January 2020, the LHCb collaboration made public a first determination of $|V_{cb}|$ from
$\bar{B}_s\to D_s^{(*)}\mu^-\bar\nu$ decays \cite{Aaij:2020hsi}.
Unfortunately, this analysis does not include the required information to repeat this determination within our framework.
However, we can determine the compatibility of the LHCb results for the quark flavour ratios $R$ and $R^*$,
\begin{equation}
    R^{(*)} \equiv \frac{\mathcal{B}(\bar{B}_s\to D_s^{(*)}\mu^-\bar\nu)}{\mathcal{B}(\bar{B}\to D^{(*)}\mu^-\bar\nu)}\,,
\end{equation}
with our theory predictions. We obtain $R = 1.038 \pm 0.034$ and $R^* = 0.975 \pm 0.076$ with a correlation
of $-2.58\%$. Our results are compatible with the LHCb results at $0.1\,\sigma$. Additionally, LHCb has published
the ratio of the $D_s^*$ branching ratio over the $D_s$ branching ratio in the semimuonic decay. Our result for
this ratio, $0.424 \pm 0.040$ is compatible with the LHCb result at $0.7\sigma$.

\acknowledgments

We thank Donal Hill for helpful discussions on the prospects of measuring the angular observables at LHCb.
We are grateful to Christoph Bobeth for pointing out a typo in the manuscript.\\

The work of MB is supported by the Deutsche Forschungsgemeinschaft (DFG, German Research Foundation) under grant 396021762 - TRR 257.
The work of MJ is supported by the Italian Ministry of Research (MIUR) under grant PRIN 20172LNEEZ.
The work of NG and DvD is supported by the DFG within the Emmy Noether Programme under grant DY130/1-1 and
the DFG Collaborative Research Center 110 “Symmetries and the Emergence of Structure in QCD”.
We thank the DFG Excellence Cluster ``Origins and Structure of the Universe'' for supporting a short-term visit of MB.

\appendix

\section{QCD Three-Point Sum Rule Results: Inputs and Results}
\label{app:QCDSRs}

\begin{table}[t]
    \centering
    \begin{tabular}{c c c c}\toprule
        parameter & value & unit & reference \\
        \midrule
         $\langle 0|\bar q q|0\rangle(2~{\rm GeV})$ & $-(0.272\pm0.010)^3$ & ${\rm GeV}^3$ & \cite{Aoki:2019cca,McNeile:2012xh,Davies:2018hmw}\\
         $R_{\bar q q}\equiv \langle 0|\bar s s|0\rangle/\langle 0|\bar q q|0\rangle$  & $1.05\pm 0.20$ & --- & \cite{McNeile:2012xh,Davies:2018hmw,Reinders:1984sr}\\
         $\left\langle 0\left|\frac{\alpha_s}{\pi}GG\right|0\right\rangle$ & $0.012\pm0.012$ & ${\rm GeV}^4$ & \cite{Shifman:1978bx,Shifman:1978by,Ioffe:2002ee,Horsley:2012ra,Bali:2014sja}\\
         $m_0^2$ & $0.8\pm0.2$ & ${\rm GeV}^2$ &  \cite{Belyaev:1982sa}\\
         $R_{\bar qGq}\equiv\langle 0|\bar s \sigma G s|0\rangle/\langle 0|\bar q \sigma G q|0\rangle$ &  $0.85\pm 0.10$ & --- & \cite{Beneke:1992ba,Braun:2004vf}\\
         \bottomrule
    \end{tabular}
    \caption{Central values and uncertainties used in the QCDSR analysis.}
    \label{tab:QCDSRs:inputs}
\end{table}

\begin{table}[t]
    \centering
    \begin{tabular}{c c}\toprule
        subleading IW & value \\
        \midrule
        $\hat\eta(1)$ & $+0.71 \, [+0.49,+0.93]$ \\[2pt]
        $\hat\eta^\prime(1)$  &   $-0.06 \, [-0.40,+0.28]$\\[2pt]
        $\hat\chi_2(1)$ &  $ -0.06 \, [-0.10,-0.02]$\\[2pt]
        $\hat\chi_2^{\prime}(1)$ & $ -0.01 \, [-0.05,+0.03]$\\[2pt]
        $\hat\chi_3^{\prime}(1)$ & $ +0.04 \, [+0.00,+0.08]$ \\[2pt]
        $\hat\eta^{(s)}(1)$ & $+0.75 \, [+0.49,+1.01]$ \\
        $\hat\eta^{(s)\prime}(1)$  &   $-0.05 \, [-0.40,+0.32]$\\
        $\hat\chi_{2}^{(s)}(1)$ &  $ -0.07 \, [-0.11,-0.03]$\\
        $\hat\chi_{2}^{(s){\prime}}(1)$ & $ +0.00 \, [-0.04,+0.04]$\\
        $\hat\chi_3^{(s)\prime}(1)$ & $ -0.01 \, [+0.03,+0.07]$\\
        \bottomrule
    \end{tabular}
    \caption{Numerical results for the subleading IW functions estimated by QCD sum-rules for both $q=u,d$ and $q=s$ cases.}
    \label{tab:QCDSRs:results}
\end{table}

We evaluate the existing three-point QCD sum rule calculations for the subleading IW functions of refs.~\cite{Neubert:1992wq,Neubert:1992pn,Ligeti:1993hw} for $q=s$ and, in order to remain consistent within our analysis, also for $q=u,d$. This generalization is possible since potential unknown perturbative or power corrections $\sim m_s$ are suppressed additionally at least by $\alpha_s \eps_c$ and included in our treatment of the uncertainties. The sum rules depend on perturbative parameters ($\alpha_s$, $\mu$), parameters pertaining only to the sum rules (Borel parameters, threshold parameters), and non-perturbative inputs (QCD condensates). The value for $\alpha_s$ is chosen consistent with the rest of our calculation, and the sum-rule specific parameters are chosen within the ranges of the original calculations. The values for the condensates are listed in \reftab{QCDSRs:inputs}. A couple of comments are in order:
\begin{itemize}
    \item We increase the uncertainty for the light-quark condensate in order to ensure consistency with the values obtained in the calculations used for the strange-quark condensate.
    \item The gluon condensate remains very difficult to calculate in general, and existing lattice calculations yield also large
    ranges. We use the ``traditional'' value \cite{Shifman:1978bx,Shifman:1978by}, but increase its uncertainty to account for other results in the literature, for instance \cite{Ioffe:2002ee,Horsley:2012ra,Bali:2014sja}.\footnote{Note that a different definition is also common in the literature, without the factor of $\pi$ in the denominator.}
    \item The parameter $m_0^2$ for the mixed quark-gluon condensate is defined via $\langle 0|\bar q \sigma G q|0\rangle\equiv m_0^2 \langle0|\bar qq|0\rangle$; its $SU(3)$-breaking seems to be under control \cite{Beneke:1992ba,Braun:2004vf}.
\end{itemize}
The $SU(3)$-breaking parameters are of the expected order; the subleading IW functions for $\bar B_s^{(*)}\to D_s^{(*)}$ are consequently compatible with the non-strange ones. 
We reproduce the central values previously obtained in refs.~\cite{Neubert:1992wq,Neubert:1992pn,Ligeti:1993hw,Bernlochner:2017jka} for $q=u,d$ when using their input values. In refs.~\cite{Ligeti:1993hw,Bernlochner:2017jka} the uncertainties for  $\eta^{(\prime)}(1)$ have been approximately doubled compared to the parametric ones, in order to account for the uncertainties inherent to the method. We follow the same recipe for the other parameters as well. We obtain the final central values and uncertainties the following way: we consider each sum rule separately, varying the sum-rule specific parameters freely within their ranges (corresponding to the R-fit treatment \cite{Hocker:2001xe}), while assuming Gaussian uncertainties for the condensates. We symmetrize the obtained interval for $\Delta\chi^2=1$ and then double the corresponding uncertainty. We do not include the resulting sizable correlations between the IW parameters, which we consider to be a conservative approach. We checked that the value obtained for $\xi^{(s)\prime}(1)$ from the sum rule for the leading IW function (which is not used as a separate theory input) is compatible with the value obtained in our fits. This is not true for the second derivative, which, however, does not enter the results for the parameters of the subleading IW functions up to the first derivative; hence we do not consider the sum-rule results for the second derivatives. Our results are summarized in \reftab{QCDSRs:results}.

\section{Light-Cone Sum Rule: Inputs and Results}
\label{app:LCSRs}

\begin{table}[t]
    \renewcommand{\arraystretch}{1.1}
    \begin{tabular}{l c c c}
        \toprule
        parameter &
            value                      &
            unit                       &
            reference                  \\
        \toprule
        $\overline{m}_c(\overline{m}_c)$ &
            $ 1.28\pm0.03$               &
            \GeV                         &
            as in \cite{Gubernari:2018wyi}                      \\
        \midrule
        $f_{B_s}$                        &
            $ 0.2307 \pm 0.0013$         &
            \GeV                         &
            \cite{Bazavov:2017lyh}       \\
        \midrule
        $f_{D_s}$                        &
            $ 0.2499 \pm 0.0004$         &
            \GeV                         &
            \cite{Bazavov:2017lyh}       \\
        \midrule
        $f_{D_s^*}$                      &
            $ 0.293 \pm 0.019$           &
            \GeV                         &
            \cite{Gelhausen:2013wia}     \\
        \midrule
        $M^2$                            &
            $ 4.5\pm 1.5$                &
            $\GeV^2$                     &
            \cite{Faller:2008tr}         \\
        \bottomrule
    \end{tabular}
    \renewcommand{\arraystretch}{1}
    \caption{%
        The central values and prior ranges for the  the charm quark mass, the decay constants and the Borel parameter used to estimate the LCSRs for the $\bar{B}_s\to D_s^{(*)}$ form factors.
    }
    \label{tab:LCSRs:inputs}
\end{table}

To obtain numerical results for the $\bar{B}_s\to D_s^{(*)}$ form factors (FFs) at and above maximum hadronic recoil, we employ the QCD sum rules on the light
cone (LCSRs) with $B$-meson distribution amplitudes (LCDAs) \cite{Faller:2008tr}. An advantageous feature of LCSRs is that the analytical form of the
results depends only on the Dirac structure of the correlator used to compute them and on universal hadronic input in form of the LCDAs, but not on the
quark flavour. We can employ the results derived in ref.~\cite{Gubernari:2018wyi} for the $B\to \{P,V\}$ transitions to compute the $\bar{B}_s\to D_s^{(*)}$
form factors to twist-four accuracy and at leading order in $\alpha_s$.

The parameters that enter the LCSRs are:
\begin{itemize}
    \item the charm quark mass $\overline{m}_c(\overline{m}_c)$ in the $\overline{\text{MS}}$ scheme;
    \item the decay constants $f_{B_s}$ and $f_{D^{(*)}_s}$ of the respective meson states;
    \item the $B_s$-to-vacuum matrix elements of local $\bar{s} G b$ currents $\lambda_{B_s,E}^2,\, \lambda_{B_s,E}^2$;
    \item the inverse moment of the $B_s$ light-cone distribution amplitude $1/\lambda_{B_s,+}$;
    \item the Borel parameter $M^2$;
    \item the duality thresholds $s_0^{(F)}$, where $F$ enumerate all of  the $\bar{B}_s\to D^{(*)}_s$ form factors.
\end{itemize}
The central values and prior ranges for the charm quark mass, the decay constants and the Borel parameter are compiled in \reftab{LCSRs:inputs} together with their
respective sources. The values used for the $\bar{s} G b$ matrix elements and the inverse moment of the $B_s$ light-cone distribution
amplitude require some comments: The $\bar{s} G b$ matrix elements provide the normalization of the three-parton LCDAs.
Their contributions to the form factors is small compared to the numerically leading two-parton terms~\cite{Gubernari:2018wyi}.
Consequently, potential $SU(3)_F$ symmetry-breaking effects are not relevant here, and we use the strict $SU(3)_F$ limit \cite{Nishikawa:2011qk}:
\begin{align}
    \lambda_{B_s,E}^2 = \lambda_{B_d,E}^2
        & = 0.03 \pm 0.02\,, &
    \lambda_{B_s,H}^2 = \lambda_{B_d,H}^2
        & = 0.06 \pm 0.03\,.
\end{align}
On the other hand, the inverse moment $1/\lambda_{B_s,+}$ of the leading-twist $B_s$ LCDA $\phi_+$ requires a more detailed discussion,
due to its bigger impact on the numerical results. To leading order in $\alpha_s$ and within the exponential model used here, 
the following relation holds\cite{Grozin:1996pq}:
\begin{align}
    \lambda_{B_q,+} = \frac{2}{3} \bar\Lambda_q\,.
    \label{eq:lambdaB}
\end{align}
However, this relation is known to be subject to UV-divergent corrections in fixed-order perturbation theory~\cite{Lee:2005gza}.
We therefore suggest to estimate the difference of $\lambda_{B_d,+}$ and $\lambda_{B_s,+}$ in which these
UV-divergent terms cancel in the $SU(3)_F$ limit.
Using $SU(3)_F$ symmetry for the power-suppressed term $\lambda_1 = -0.30\,\GeV^2$,
the forward matrix element of the kinetic operator, we obtain
$\bar\Lambda_d = 0.500\,\GeV$ and $\bar\Lambda_s = 0.590\,\GeV$.
To be consistent with the previous LCSR analysis of the $\bar{B}\to D^{(*)}$ form factors \cite{Gubernari:2018wyi} we
use $\lambda_{B_d,+} = 0.460\pm 0.110\,\GeV$ \cite{Braun:2003wx}, and estimate:
\begin{align}
    \label{eq:lambdaBs}
    \lambda_{B_s,+}
        & = \lambda_{B_d,+} + \frac{2}{3}\left(\bar\Lambda_s - \bar\Lambda_d\right)
          = 0.520 \pm 0.110\,\GeV\,.
\end{align}

For our analysis we adopt the same Borel parameters as for the $\bar{B}\to D^{(*)}$ analysis carried out in ref~\cite{Gubernari:2018wyi}.
We also ensure that
\begin{enumerate}
    \item under variation of the Borel parameters $M^2$ in the chosen intervals the sum rule yields stable results;
    \item the contributions due to continuum and excited states above the respective thresholds $s_0^{(F)}$ are small compared to the ground state contribution,
    i.e., the integral from $s=0$ to $s_0^{(F)}$;
    \item contributions at higher twists remain small.
\end{enumerate}
The variation of our sum rules in the Borel windows given in \reftab{LCSRs:inputs} contributes $9\%$ to the overall
systematic uncertainty of our results, which is larger than what was obtained for the $\bar{B}\to D^{(*)}$ analysis~\cite{Gubernari:2018wyi}.
We further account for the absence of $1/m_b^2$ in the correlator by assigning an additional $5\%$ to the systematic uncertainty.
Adding the two in quadrature yields an overall systematic uncertainty of $\sim 10\%$.

The thresholds $s_0^{(F)}$ are determined using the same procedure as proposed in ref.~\cite{Imsong:2014oqa} and subsequently employed
in ref.~\cite{Gubernari:2018wyi}. The basic idea is to take the derivative of the FF sum rule with respect to $-1/M^2$ and to normalize
the derivative to the FF sum rule itself, obtaining (schematically) the squared meson mass-estimator
\begin{align}
    \label{eq:daughterSR}
    \left[M_{D^{(*)}}^2\right]_{\text{LCSR}} = \frac{\int_0^{s_0^{(F)}}e^{-\frac{s}{M^2}} s\, \rho_F(s,q^2)}{\int_0^{s_0^{(F)}}e^{-\frac{s}{M^2}} \, \rho_F(s,q^2)}\,,
\end{align}
with $\rho_F$ standing in for the spectral density from which we extract the form factor $F$.
Following~\cite{Gubernari:2018wyi}, we impose $5\%$ uncertainties on the estimator of the squared meson mass, to account for higher
twist correction to the spectral density $\rho_F(s,q^2)$. We also vary $q^2$ from $-15~\GeV^2$ to $0~\GeV^2$, for which we find
that $q^2$ dependence of the estimator $[M^2]_\text{LCSR}$ is negligible. We use $s_0^{(F)}(q^2) \equiv s_0^{(F)}$.
The union of the threshold intervals at $68\%$ probability for each of the FFs\footnote{%
    Except for $f_T^{B\to D}$, for which the threshold determination has not been possible for the same reasons illustrated in~\cite{Gubernari:2018wyi}.
    We follow the same procedure as outlined there to estimate the $f_T$ threshold parameter using the $f_+$ threshold parameter.
} reads:
\begin{align}
    s_0^{(F)} & = [6.9,~11.0]\,\GeV^2  & & \text{ for } \bar{B}_s\to D_s\,,  \\
    s_0^{(F)} & = [7.9,~11.8]\,\GeV^2  & & \text{ for } \bar{B}_s\to D_s^*\,.  
\end{align}
Our predictions for the full set of form factors and for the set excluding $f_T$, both evaluated at $q^2=\{-15,-10,-5,0\}~\GeV^2$
and including the covariance matrix across form factors and across $q^2$ points, are published as part of the \EOS software \cite{EOS} as of version v0.3.1.
Both sets of predictions can be accessed as multivariate Gaussian constraints named
\begin{center}
    \verb|B_s->D_s^(*)::FormFactors[f_+,f_0,f_T,A_0,A_1,A_2,V,T_1,T_2,T_23]@BGJvD2019|,\\
    \verb|B_s->D_s^(*)::FormFactors[f_+,f_0,A_0,A_1,A_2,V,T_1,T_2,T_23]@BGJvD2019|,
\end{center}
respectively. In addition, we provide these predictions as machine readable ancillary files attached to this preprint.

\bibliographystyle{apsrev4-1}
\bibliography{references}

\end{document}